\documentclass[]{article}
\usepackage{comment}
\usepackage{graphicx}        
\usepackage{multicol}        

\usepackage[usenames,dvipsnames,svgnames,table]{xcolor}
\usepackage[margin=1.0in]{geometry}

\usepackage{color}
\usepackage{hyperref}
\usepackage{amsmath,amssymb}
\graphicspath{{./img/}}

\begin{document}

\title{A Pulsed-Precipitation Model of Dryland Vegetation Pattern Formation}
\author{
	Punit Gandhi\thanks{Department of Mathematics and Applied Mathematics, Virginia Commonwealth University, Richmond, VA 23284, USA}
	, Lily Liu\thanks{University of Chicago, Chicago, IL 60637, USA}
	and Mary Silber\thanks{Department of Statistics and the Committee on Computational and Applied Mathematics,  University of Chicago, Chicago, IL 60637, USA}
}	
	\maketitle
	
	\begin{abstract}
We develop a model for investigating the impact of rainstorm variability on the formation of banded vegetation patterns in dryland ecosystems. Water input, during rare rainstorms, is modeled as an instantaneous kick to the soil water. The redistribution, from surface water to soil moisture, accounts for the impact of vegetation on infiltration rate and downslope overland flow speed. These two positive feedbacks between water and biomass distributions act on the fast timescales of rain storms. During dry periods, a classic reaction-diffusion framework is used for the slow processes associated with soil water and biomass. This pulsed precipitation model predicts that the preferred spacing of the vegetation bands is determined by the characteristic distance that a storm pulse of water travels overland before infiltrating into the soil. In this way, the vegetation pattern is determined by the fast ecohydrological processes and may be attuned with its dryland precipitation pattern. We demonstrate how this modeling framework, suited for stochastic rain inputs, can be used to investigate possible collapse of a dryland pattern-forming ecosystem under different precipitation patterns with identical low annual mean. Model simulations suggest, for instance, that shorter rainy seasons and greater variability in storm depth may both hasten ecosystem collapse.

	\end{abstract}

\section{Introduction}
\begin{figure}
    \centering
    \includegraphics[width=\textwidth]{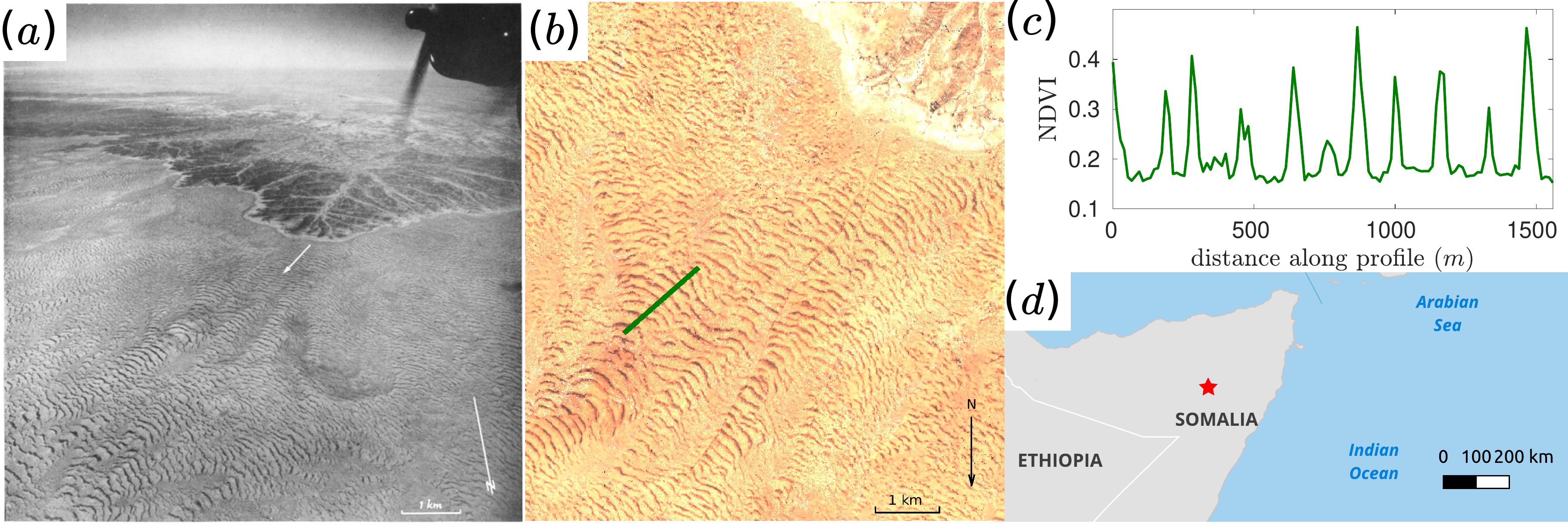}
    \caption{(a) Aerial photograph of banded vegetation patterns  in the Sool region of Somalia, taken in March 1945~\cite{macfadyen1950vegetation,gowda2018signatures}. (b) Satellite image  of the same vegetation pattern taken in October 2020 from the Copernicus Sentinel 2 mission~\cite{DRUSCH201225}. (c) Normalized difference vegetation index (NDVI) along green line shown in panel (b).  The location of the pattern (9$^\circ$20'38"N 48$^\circ$46'22"E) is indicated on the map  in panel (d). }
    \label{fig:intro}
\end{figure}

The availability of aerial photography in the 1940's first enabled the study of  landscape-scale spatial patterns of vegetation growth in the Horn of Africa~\cite{macfadyen1950soil,macfadyen1950vegetation}.  On gentle slopes, $\lesssim 2\%$ grade, the patterns typically consist of bands of dense vegetation that are tens of meters wide and separated by bare soil (Figure~\ref{fig:intro}). They exhibit regular spacing, with wavelength on the order of a hundred meters, and are oriented approximately perpendicular to the elevation grade. More recent studies incorporating modern satellite images have reported little change, relative to initial aerial photographs, at least in absence of increased human pressure~\cite{gowda2018signatures}; the most remarkable change is a slow uphill migration of the pattern, on order of meters per decade~\cite{deblauwe2012determinants}.  It is now known that spontaneous formation of periodic vegetation patterns occurs in  drylands around the globe~\cite{deblauwe2008global}. Mathematical models suggest that the phenomenon may be a strategy to exploit positive feedbacks that concentrate, in the vegetated zones, the limiting water resource~\cite{borgogno2009,meron2015nonlinear,gandhi2019vegetation}. 

The striking regularity of the dryland vegetation patterns has led to proposals that they may possess remotely-sensed characteristics that are indicative of the ecosystem health and its risk of collapse~\cite{rietkerk2004self,dakos2011slowing}. If true, monitoring changes in the patterns over time could provide information about the resilience of the ecosystems that support them, including their vulnerability under climate change.
Many of the mathematical modeling studies have focused on pattern transitions, using  mean annual precipitation level as a bifurcation parameter~\cite{von2001diversity,gilad2004ecosystem,rietkerk2004self,gowda2014transitions,gowda2016assessing}. Indeed there is some observational evidence of significant changes in pattern morphology along an aridity gradient from South Sudan into Sudan~\cite{deblauwe2011environmental}.
Here we aim to expand the use of mathematical models to investigate changes in patterns, including possible collapse, under other characteristics of dryland precipitation, such as variability in storm frequency, storm depth, and  length of rainy seasons. 

Much of the currently available spatial remote sensing data provides information about the biomass distribution which, while fluctuating on a seasonal timescale,  evolves on a decade timescale or longer. However, feedbacks between water and vegetation that are thought to be responsible for pattern formation involve a much faster rainstorm timescale.  Detailed mechanistic models that attempt to bridge these disparate timescales can require significant computational resources to make useful predictions on the timescale of pattern evolution~\cite{paschalis2016matching,fatichi2012mechanistic}. Our approach to this challenge builds on  a fast-slow switching framework~\cite{gandhi2020fast} developed specifically to capture the processes involved, at a conceptual level, using the range of timescales on which they occur. We make some further simplifying assumptions about the fast processes that lead to a computationally tractable model for carrying out the large number of trials required to explore the impact of rainfall variability.

In contrast to the fast-slow switching model developed in~\cite{gandhi2020fast}, we do not attempt to resolve the short intra-rainstorm time in this current work, and  instead treat each rain event as a Dirac-delta function impulse that deposits a uniform layer of water on the surface.  We convert this surface water layer directly into an increase in the soil water distribution, taking into account the key feedbacks between the biomass distribution and (1) overland flow speed, (2) infiltration rate.  
In analogy with ``flow-kick'' systems considered in the context of ecological resilience~\cite{meyer2018quantifying}, the rain events become instantaneous ``kicks" to the soil water in the reaction-diffusion model that governs the ``flow", or time-evolution, of the ecosystem via water-biomass interactions. We use the resulting impulsive reaction-diffusion system~\cite{lewis2012spreading} with nonlocal, spatially heterogeneous impulses to investigate vegetation bands on a one-dimensional hillslope with stochastic rainfall. 

Many  of the earliest conceptual PDE modeling efforts handle the multiple scales associated with vegetation pattern formation by formulating a model on an annually averaged timescale in which the fast processes are phenomenologically ``upscaled''~\cite{Klausmeier1999,rietkerk2002self,gilad2004ecosystem}. Mathematical analysis of such models that highlighted the mechanisms that set the spacing between vegetation bands or patches  formulated this in terms of the wavenumber of Turing patterns in the context of reaction-diffusion models, or the analogous Turing-Hopf patterns when advection is also present~\cite{sherratt2005analysis,van2013rise,bastiaansen2018multistability}.  A key feature of these models is that the wavenumber, and thus the predicted band or patch spacing, is controlled by the relative strengths of the transport terms, e.g. the phenomenological diffusion and advection constants incorporated into surface water, soil water and biomass equations. 
Our analysis of the proposed pulsed-precipitation model has the characteristic band spacing set instead by the typical storm depth, which controls the distance that deposited surface water travels before it infiltrates into the soil. Moreover, the associated wavenumber is largely independent of the phenomenological diffusion constants incorporated into the soil water and biomass equations. In this way our results deviate from those of earlier conceptual models, and suggest interesting directions for future studies of more mechanistically detailed models and for field studies of the overland flow and infiltration that takes place after a storm, e.g. of the type reported in~\cite{cornet1988dynamics}. An interesting point of comparison can already be made between our wavenumber selection results and those presented in a recent paper by Crompton and Thompson~\cite{crompton2021sensitivity}. Their study uses a different approach to determining the soil moisture distribution following a storm event. It is based on using machine learning to build an emulator for the Saint Venant shallow water equations for overland flow, coupled to the Richards equation for infiltration.
They also find that greater storm depth, with identical annual mean, leads to increased spacing of the bands.

The impact of rainfall seasonality and variability on vegetation pattern dynamics is a major motivation for our study, and has been explored in other works~\cite{dordorico2006vegetation,guttal2007self,kletter2009patterned,siteur2014will,eigentler2020effects}.
Here, we incorporate into the pulsed-precipitation model a stochastic rainfall that  assumes a Poisson point process for storm arrivals and draws each storm depth from an exponential distribution~\cite{rodriguez1999probabilistic}. We demonstrate how this model can be used to investigate the likelihood of noise-induced transitions between patterned states and  barren desert state in a low annual-mean-rainfall, bistable regime.
Specifically, we find that the vegetation can spontaneously collapse in a finite time as a result of fluctuations in rainfall and that both precipitation characteristics, such as the mean storm depth, and pattern characteristics, such as the band spacing,  
impact how long the vegetation survives on average. We find, for example, that greater variability in storm depth increases the likelihood of collapse. The model also predicts that the same stochastic rainfall pattern, spread out over a longer rainy season, leads to longer-lived vegetation. 

Our paper is organized as follows.  In Section~\ref{sec:fastslow}, we summarize the fast-slow switching framework~\cite{gandhi2020fast}  that serves as a foundation for the pulsed-precipitation model used in this study,  highlighting the key simplifications  that  make our stochastic rainfall simulations possible.  In Section~\ref{sec:pulsedprecip}, we leverage these simplifying assumptions to derive a pulsed-precipitation model in which rainstorms act as kicks to the soil moisture, which then evolves slowly, with the biomass, during the long dry periods between storms. In Section~\ref{sec:per}, we investigate pattern-forming instabilities  of  spatially uniform states of the pulsed-precipitation model for an idealized periodic sequence of rain pulses. This linear stability analysis reveals a spatial resonance tongue structure that suggests the distance that surface water travels before infiltrating into the soil plays a key role in wavelength selection.  In Section~\ref{sec:stoch}, we numerically explore important qualitative differences in the dynamics of the model under stochastic rainfall versus periodic rainfall, while also showing that the wavelength, as in the linear problem, is tuned to mean overland flow distance of surface water following a storm.  We then demonstrate how the model can be used to probe possible ecosystem collapse that results from rainfall variability in the bistable regime. Finally, in Section~\ref{sec:dis}, we discuss the results of our study in the context of other related work, and suggest potential directions to pursue with the pulsed-precipitation model. 

\section{Fast-Slow Switching Framework}
\label{sec:fastslow}

This section introduces a model for the formation of banded vegetation patterns based on the fast-slow switching framework developed in~\cite{gandhi2020fast}.   The switching framework evolves, on appropriate timescales, three fields: surface water height $H(X,T)$ [$cm$], soil water column height $W(X,T)$ [$cm$], and biomass density $B(X,T)$ [$kg/m^2$]. While $H(X,T)$ only evolves on the short time scale of rain events and $B(X,T)$ only evolves on the long time scales between them, $W(X,T)$ responds to processes that act on the fast timescale and other ones that occur on the slow timescale. 

The output of the fast part of the switching model, after surface water has infiltrated the soil under an assumption of  fixed biomass distribution ${\cal B}(X)$, is a soil moisture distribution ${\cal W}(X)$. This is the initial condition for the slow system that applies during the 
ensuing dry period between rain storms.
The slow system evolves both biomass and soil moisture. It takes into account evapo-transpiration of  soil moisture, biomass growth and death, as well as seed dispersal, modeled as biomass diffusion, which leads to up-slope colonization of the vegetation. The explicit formulation of the fast and slow parts of the switching model are given in Sections~\ref{sec:fastslow:fast} and~\ref{sec:fastslow:slow}, respectively.  We highlight the differences between the original formulation in~\cite{gandhi2020fast} and the version used in this paper.

The modifications we introduce here allow for a closed form solution for the spatial distribution of soil water ${\cal W}$ once all of the surface water from a rainstorm  has infiltrated into the soil. In Section~\ref{sec:pulsedprecip}, we leverage this result to formulate a pulsed-precipitation model in which the biomass $B(X,T)$ and soil water $W(X,T)$ evolve on the slow timescale and rainstorms are treated as instantaneous impulses to $W(X,T)$, determined by the closed form solution of the fast system, for a given storm  depth and the current biomass profile. 

Both the switching model and  the pulsed-precipitation model are formulated on a one-dimensional spatial domain with periodic boundary conditions; the  perspective is that it is capturing some representative middle portion of a long swath of gently sloped terrain, oriented with uphill in the $+X$ direction. 
We conclude this section in~\ref{sec:fastslow:rainfall} by presenting the rainfall models used in this study, inspired by a typical climatology in the Horn of Africa.

\subsection{Fast Subsytem of the Switching Model}
\label{sec:fastslow:fast}
The fast portion of the switching model is
\begin{subequations}
	\label{equationswitch1}
\begin{align}
\frac{\partial H}{\partial T}&=\underbrace{P(T)}_{Precip.}
-\underbrace{{\cal I}
\Bigl(H,W;{\cal B}(X)\Bigr)}
_{Infiltration}
+\underbrace{\frac{\partial}
{\partial X}\Bigl({\cal V}(H;{\cal B}(X))\ H\Bigr)}
_{Advection}
\label{eqH1} \\
\frac{\partial W}{\partial T}&=
\underbrace{{\cal I}
\Bigl(H,W;{\cal B}(X)\Bigr)}
_{Infiltration}
,\label{eqW1}
\end{align} 
\end{subequations}
where the infiltration rate, ${\cal I}$ $[cm/day]$, and overland surface flow speed, ${\cal V}$ $[m/day]$, are given in Table~\ref{tab:infadv} for both the original fast slow model and the  approximation that leads to the pulsed-precipitation model.

\begin{table}[tbp]
	\renewcommand{\arraystretch}{2.55}
	\centering
	\everymath{\displaystyle}
	\begin{tabular}{|c|c|c|c|} 
	    \hline
	    & Fast-Slow 
	    ~\cite{gandhi2020fast}
	    & Pulsed-Precipitation 
	    \\
		\hline
		 Infiltration Rate 
		 &$\overline{K}_I\Bigl(\frac{{\cal B}(X)+fQ}{{\cal B}(X)+Q}\Bigr)\Bigl(\frac{H}{H+A} \Bigr)\Bigl(1-\frac{W}{W_s}\Bigr)^{\beta_I}$
		 & $K_I\Bigl(\frac{{\cal B}(X)+fQ}{{\cal B}(X)+Q}\Bigr)\Theta(H)$\\
		 \hline
		 Surface Flow Speed 
		 &  $ \Bigl(\frac{ K_V\sqrt{\zeta} }{1+N{\cal B}(X)}\Bigr) H^{\beta_V} $
		 & $\frac{ K_V\sqrt{\zeta} }{1+N{\cal B}(X)}$
		 \\
		 \hline
	\end{tabular}
	\caption{\label{tab:infadv} Comparison of infiltration rate $\mathcal{I}$ and overland water flow speed $\mathcal{V}$ used with fast-slow switching model in~\cite{gandhi2020fast} and our modified 
	pulsed-precipitation model. For the pulsed-precipitation model, we set $K_I=200\, cm/day$, whereas for the fast-slow model, we used $\overline{K}_I=500\, cm/day$ with $A=1\,cm$, $W_s=27\,cm$, and $\beta_I=4$ for the additional factors that can reduce infiltration. Both $\beta_V=2/3$ (with $K_V\sqrt{\zeta}= 1.4\, m/day/cm^{2/3}$) and the computationally-faster $\beta_V=0$ (with $K_V\sqrt{\zeta}= 1.4\, m/day$) were used in~\cite{gandhi2020fast}.  All other parameters match values for the pulsed-precipitation model given in Table~\ref{tab:dim}.} Here $\Theta(H)$ denotes a Heaviside step function in $H$. 
	\end{table}
In the pulsed model, the $H$ and $W$ dependent factors in ${\cal I}$ are replaced by a Heaviside step function in $H$; infiltration occurs whenever there is water on the surface at a rate that depends only on ${\cal B}$ at that location. 
With the default value of $f=0.1$, the bare soil infiltration rate (${\cal B}=0$ $kg/m^2$) is a factor of 10 slower than its maximum rate for ${\cal B}\gg Q$.
This sigmoidal transition from low to high infiltration rate with increasing biomass ${\cal B}$ is an essential positive feedback, and suggests an advantage for ${\cal B}$ to exceed the threshold $Q$ in its patterned state. 

We assume a constant $0.5\%$ elevation grade ($\zeta=0.005$) and, as is common in conceptual models for vegetation pattern formation, we omit the dependence on $H$ in the speed of overland flow ${\cal V}$ for the pulsed-precipitation model. We note that exactly what dependence is most appropriate in this setting is still an open question, and the impact on qualitative predictions may potentially be minimized by appropriately calibrating the model~\cite{crompton2020resistance}. 
The overland flow speed is decreased, by a factor $1+N{\cal B}(X)$,  if there is  vegetation at a location $X$. 
This longer residence time, of surface water on vegetated soil compared to bare soil, is another positive feedback between water resource and biomass that acts on the fast timescale.

\subsection{Slow Subsystem of the Switching Model}
\label{sec:fastslow:slow}
The slow portion of the switching model evolves the  soil water $W(X,T)$,  initialized by its post-storm distribution ${\cal W}(X)$, and the biomass density $B(X,T)$. Specifically,
\begin{subequations}
	\label{equationswitch2}
\begin{align}
\frac{\partial W}{\partial T}&=-\underbrace{\Bigl(L+\Gamma B\Bigr)W}_{Evapotranspiration}+\underbrace{D_W\frac{\partial^2 W}{\partial X^2}}_{Diffusion}\label{eqW2}\\
\frac{\partial B}{\partial T} &= \underbrace{C\Bigl(1-\frac{B}{K_B}\Bigr)\Gamma B W}_{Growth} -\underbrace{MB}_{Death}+\underbrace{D_B\frac{\partial^2 B}{\partial X^2}}_{Dispersal}.
\label{eqB2}
\end{align} 
\end{subequations}
Here the evaporation rate is given by $L$ and the transpiration rate is given by $\Gamma B$. Transpiration dictates the biomass growth rate with an efficiency set by the parameter $C$, and with a logistic term that limits growth if $B$  approaches a carrying capacity $K_B$. The death rate $M$ is constant and seed dispersal is modeled by linear diffusion.  As is done in~\cite{gandhi2020fast}, we typically neglect the soil water diffusion, i.e. $D_W=0$. Our simulation results with $D_W>0$ indicate that soil water diffusion plays a negligible role in the model.  See Appendix~\ref{appendix:DBW} for numerical exploration of the impact of diffusion rates $D_B$ and $D_W$ on pattern formation within the pulsed-precipitation model.

\subsection{Precipitation Model}
\label{sec:fastslow:rainfall}

\begin{figure}
    \centering
    \includegraphics[width=\textwidth]{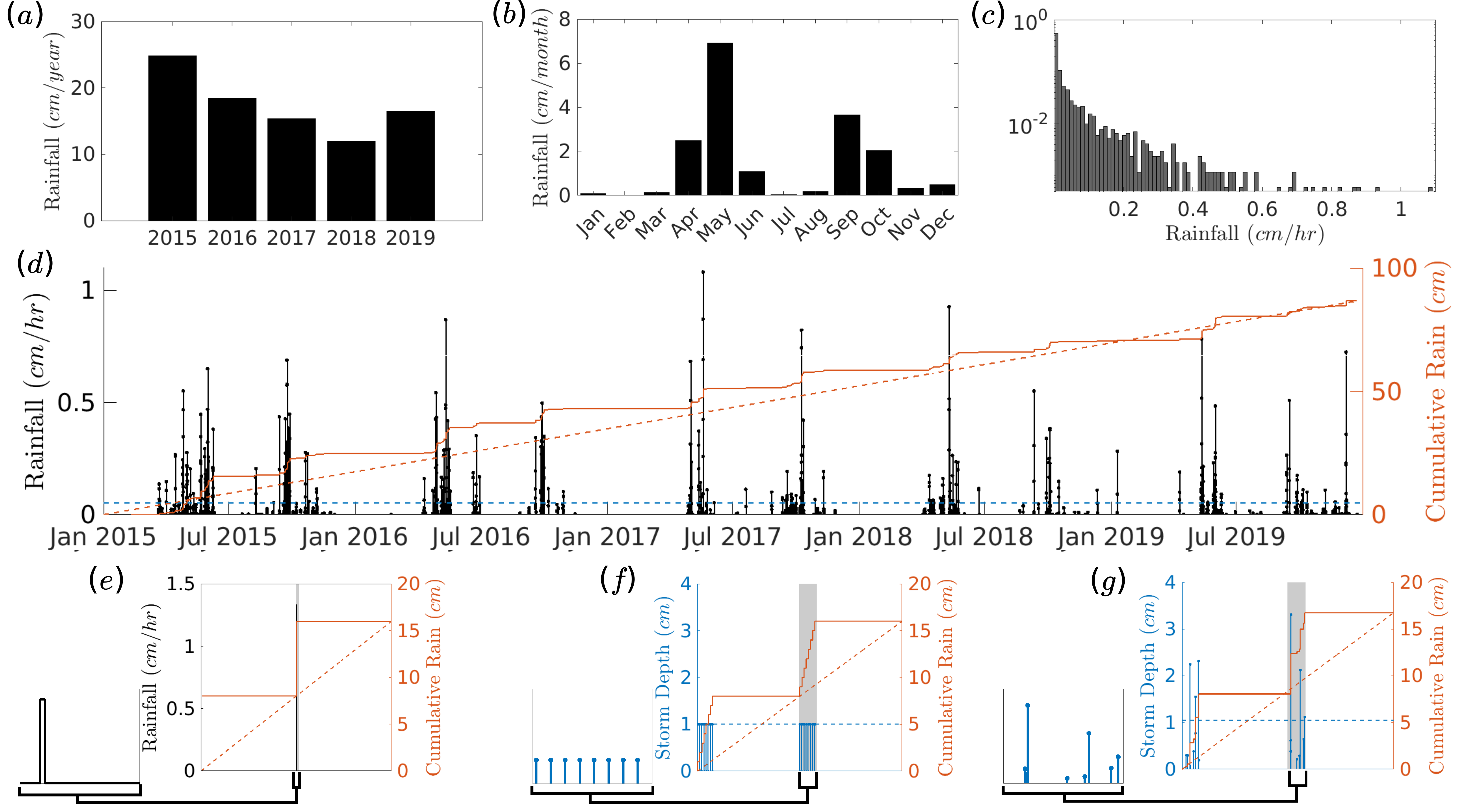}
    \caption{(a) Annual totals,  (b) Average monthly totals and (c) rainfall rate distribution for  (d) five years of half-hourly reanalysis rainfall  data~\cite{huffman2015nasa} at the location from Figure~\ref{fig:intro}(d).  Also shown is time series generated by (e) periodic rainfall model used in~\cite{gandhi2020fast}, consisting of a six-hour storm with storm depth of 8 $cm$, repeating every six months, (f) periodic rainfall model with eight instantaneous pulses with storm depth of 1 $cm$, evenly spaced in each one-month biannual rainy season, and (g) stochastic rainfall model with two one-month rainy seasons per year, mean storm depth of 1 $cm$, and mean annual precipitation of 16 $cm$.  Note that rainfall in the pulsed precipitation model is characterized by storm depth in $cm$ (blue), whereas the rainfall in the fast-slow model is given in terms of a rainfall rate $cm/hr$ over a given interval of time (black).} 
    \label{fig:rainfall}
\end{figure}

We use rainfall patterns in the Horn of Africa as inspiration for our rainfall models. Figure~\ref{fig:rainfall}(a-d) shows rainfall statistics, based on reanalysis data of rainfall rates~\cite{huffman2015nasa} at the site shown in Figure~\ref{fig:intro}, between 2015 and 2020,  along with associated cumulative rainfall.  

Rainfall statistics in Figure~\ref{fig:rainfall}(a-c), which are based on rainfall rates shown in Figure~\ref{fig:rainfall}(d), indicate two rain seasons per year, each lasting approximately 1-2 months with annual precipitation fluctuating between 11 and 25 $cm/year$ over the 5 year period.
We emphasize that the data presented in Figure~\ref{fig:rainfall}, while corrected using available rain gauge data, is reanalysis data based on models and not directly measured. We can therefore reliably report rainfall rate statistics, but not rainstorm depth statistics, which would be most relevant for informing the pulsed precipitation model.  

Investigations in~\cite{gandhi2020fast}, with the fast-slow switching model,  collapsed the rainfall of each rainy season into a single hours-long storm of constant intensity, as illustrated in Figure~\ref{fig:rainfall}(e). The pulsed-precipitation model, introduced in Section~\ref{sec:pulsedprecip}, assumes each rainstorm instantaneously deposits water on the surface.  We note that this is an assumption of convenience, and other studies have explored the role that  storm duration can play in vegetation patterns~\cite{crompton2021sensitivity}. In  Appendix~\ref{appendix:CompareFastSlow}, we show how we might capture the effects of storm duration in the pulsed-precipitation model by interpreting the storm depth as an effective surface water height during a storm.

In order to carry out linear stability analysis in Section~\ref{sec:per}, we consider a periodic array of identical, evenly spaced rain pulses within each rainy season, as illustrated in Figure~\ref{fig:rainfall}(f). In Section~\ref{sec:stoch} we explore the dynamics under a stochastic rainfall model~\cite{rodriguez1999probabilistic} that treats rainstorm arrivals as a Poisson point process, during each fixed duration rainy season, with storm depths drawn from an exponential distribution, as illustrated in Figure~\ref{fig:rainfall}(g).

\section{Pulsed-Precipitation Model}
\label{sec:pulsedprecip}
This section develops the model we use in this paper for stochastic precipitation simulations.
While we retain the same reaction-diffusion model \eqref{equationswitch2} for the slow subsystem, we make two significant changes  to the fast subsystem \eqref{equationswitch1}, which allow us to determine its output soil moisture distribution ${\cal W}(X)$ in a closed form, by quadrature. The model changes are:
\begin{enumerate}
    \item The precipitation $P(T)$ in~\eqref{eqH1} is replaced by rain events that instantaneously deposit a column of water, of height $H_0$, uniformly on the domain. The  timing and strength of these ``precipitation pulses" are the random variables in our stochastic simulations.
    \item The infiltration rate  used in~\cite{gandhi2020fast}, given in Table~\ref{tab:infadv}, is replaced  by
    \begin{equation}
        {\cal I}(H;{\cal B}(X))\equiv K_I\Bigl(\frac{{\cal B}(X)+fQ}{{\cal B}(X)+Q}\Bigr)\Theta(H).
    \end{equation}
    The infiltration is independent of how saturated the soil might be, and has a simple on-off switch with presence-absence of surface water. We model this with the Heaviside unit step function $\Theta(H)$, assuming the convention that $\Theta(0)=0$.
\end{enumerate}
With these modifications we are able to reformulate the fast-slow model into a pulsed-precipitation framework, with rain input modeled by instantaneous kicks to the soil water followed by evolution of the slow system during the intervening dry-surface time intervals. We compare results from the two models in Appendix~\ref{appendix:CompareFastSlow}. 

The non-dimensionalization
in Section~\ref{sec:pulsedprecip:dim} reveals key characteristic scales associated with the modified fast and slow subsystems~\eqref{equationswitch1}-\eqref{equationswitch2}.  In Section~\ref{sec:pulsedprecip:solve}, we solve the fast system  to obtain the spatial distribution of water that has infiltrated  the soil following a Dirac-delta rain impulse. Algebraic manipulation of the integral expression for the soil water kick provides a geometric interpretation of the infiltration process that redistributes water from the surface into the soil.

\subsection{Dimensionless Parameters}
\label{sec:pulsedprecip:dim}
In this subsection we present the dimensionless version of the pulsed precipitation model used in our investigations. For this, we introduce two different (dimensionless) timescales, $t$ and $\tau$ for the fast and slow subsystems, respectively, and a dimensionless distance $x$. Specifically, we let
\begin{equation}
\label{eq:nonddef1}
    t=\frac{K_I}{{\cal H}_0}T,\quad \tau=MT, \quad x=\frac{K_I/{\cal H}_0}{K_V\sqrt{\zeta}}X.
\end{equation}
Here ${\cal H}_0$ is a characteristic rain pulse height and ${\cal H}_0/K_I$ is an associated infiltration timescale. This time, together with a characteristic overland flow speed $(K_V\sqrt{\zeta})$, determines a characteristic overland travel distance  $({\cal H}_0/K_I)/(K_V\sqrt{\zeta})$ that is used to non-dimensionalize $X$.  We set the (slow) biomass timescale by its mortality rate,  $M$.  Finally, we define the dimensionless fields:
\begin{equation}
\label{eq:nonddef2}
    h=\frac{H}{{\cal H}_0},\quad w=\Bigl(\frac{C\Gamma}{M}\Bigr)W,
    \quad b=\frac{B}{Q}.
\end{equation}

\begin{table}[tbp]
	\renewcommand{\arraystretch}{1.55}
	\centering
	\begin{tabular}{|c|c|c|l|} 
		\hline
		parameter & units& default value & description/definition \\
		\hline
		\hline
		${\cal H}_0$ & $cm$ & 1 & characteristic precipitation pulse \\
		\hline
		$K_I$ & $ cm/day$ & 200 & infiltration rate coefficient  \\
		\hline
		$f$ & -- & 0.1 & bare/vegetated infiltration contrast\\
		\hline
		$Q$ & $kg/m^2$ & 0.1 & biomass level for infiltration enhancement \\
		\hline
		$K_V\sqrt{\zeta}$& $m/day$ & $1.4\times 10^4$ & surface water speed (bare soil)\\
		\hline
		$N$ & $m^2/kg$& 20 & surface roughness coefficient\\
		\hline
		$L$ & $day^{-1}$ &0.0075& evaporation rate\\
		\hline
		$\Gamma$ & $(kg/m^2)^{-1}day^{-1}$ & 0.025 & transpiration coefficient\\
		\hline
		$K_B$ & $kg/m^2$ & 4 & biomass carrying capacity\\
		\hline
		$C$ & $(kg/m^2)/cm$ &0.1 & water use efficiency coefficient\\
		\hline
		$M$& $day^{-1}$& 0.01 & biomass mortality rate\\
		\hline
			$D_B$& $m^2/day$& $0.01$ &  biomass diffusion\\
		\hline
			$D_W$& $m^2/day$& $0$  &  soil water diffusion\\
		\hline
		\hline
		$\eta$& --& 2 & $\eta\equiv NQ$\\
		\hline
		$\alpha$ & --& 0.25 & $\alpha\equiv {\cal H}_0 C\Gamma/M$\\
		\hline
		$\sigma$ &--& 0.75& $\sigma\equiv L/M$\\
		\hline
		$\gamma$ &--&0.25 & $\gamma\equiv \Gamma Q/M$\\
		\hline
		$\kappa$& -- & 40& $\kappa\equiv K_B/Q$\\
		\hline
		$\delta_w$ & -- & 0 & $\delta_w\equiv D_WK_I^2/(M {\cal H}_0^2K_V^2\zeta)$
		\\
		\hline
		$\delta_b$ & -- &  0.0002 & $\delta_b\equiv D_BK_I^2/(M{\cal H}_0^2K_V^2\zeta)$
		\\
		\hline
	\end{tabular}
	\caption{\label{tab:dim} Summary of parameters used in numerical simulations. Values are given for the dimensioned fast-slow model~\eqref{equationswitch1}-\eqref{equationswitch2} and the non-dimensionalized pulsed precipitation model \eqref{eq:fast:nondim}-\eqref{eq:slow:nondim}. We neglect soil water diffusion so $\delta_w=0$. 
	}
\end{table}%

The fast subsystem of the pulsed-precipitation model, in dimensionless variables, is
\begin{subequations}
	\label{eq:fast:nondim}
\begin{align}
\frac{\partial h}{\partial t}&= -\iota(x) \Theta(h) +\frac{\partial}{\partial x}\left( \nu(x) h
\right)\label{eq:fast:nondim:h}\\
\frac{\partial w}{\partial t} &= \alpha \iota(x) \Theta(h), \label{eq:fast:nondim:w}
\end{align} 
\end{subequations}
where
\begin{equation}
\label{eq:iotanu}
\iota(x)=\frac{\hat{b}(x)+f}{\hat{b}(x)+1},\quad  \nu(x)=\frac{1}{1+\eta \hat{b}(x)}.
\end{equation}
Because the rainstorm is assumed to deposit water on the surface instantaneously, we take $h=h_0$ as the initial condition for Equation~\eqref{eq:fast:nondim:h}, and there is no longer an explicit precipitation term. The dimensionless biomass distribution $\hat{b}(x)\equiv {\cal B}(X)/Q$ is taken from the slow subsystem at the arrival time of the precipitation pulse. The non-dimensionalized slow subsystem is
\begin{subequations}
	\label{eq:slow:nondim}
\begin{align}
\frac{\partial w}{\partial \tau}&= \delta_w\frac{\partial^2 w}{\partial x^2}-(\sigma+\gamma b) w\label{eq:slow:nondim:w}\\
\frac{\partial b}{\partial \tau} &=\delta_b \frac{\partial^2 b}{\partial x^2}+ w b\left(1- \frac{b}{\kappa}\right) - b. \label{eq:slow:nondim:b}
\end{align} 
\end{subequations}

Definitions of the dimensionless parameters, and typical values used in simulations for all dimensioned parameters of the model, are given in Table~\ref{tab:dim}; see~\cite{gandhi2020fast} for details of these parameter estimates. Note that for the default parameters, the fast infiltration time-scale in \eqref{eq:nonddef1} is less than 10 minutes, while the slow time-scale associated with the biomass is 100 days. The characteristic distance for overland water flow is $\sim 70\ m$. This contrasts with the short biomass diffusion scale of $2\sqrt{365 D_B} \approx 2.4 m$, which is based on a year timespan. The disparity of these two length-scales is reflected in the non-dimensionalized model through small diffusion parameter $\delta_b\ll 1$. Finally, we note that the characteristic soil water depth in \eqref{eq:nonddef2} is $W_0=M/C\Gamma=4 cm=4{\cal H}_0$.

\subsection{Solving the fast subsystem of the pulsed precipitation model}
\label{sec:pulsedprecip:solve}

\begin{figure}
    \centering
    \includegraphics[width=0.95\textwidth]{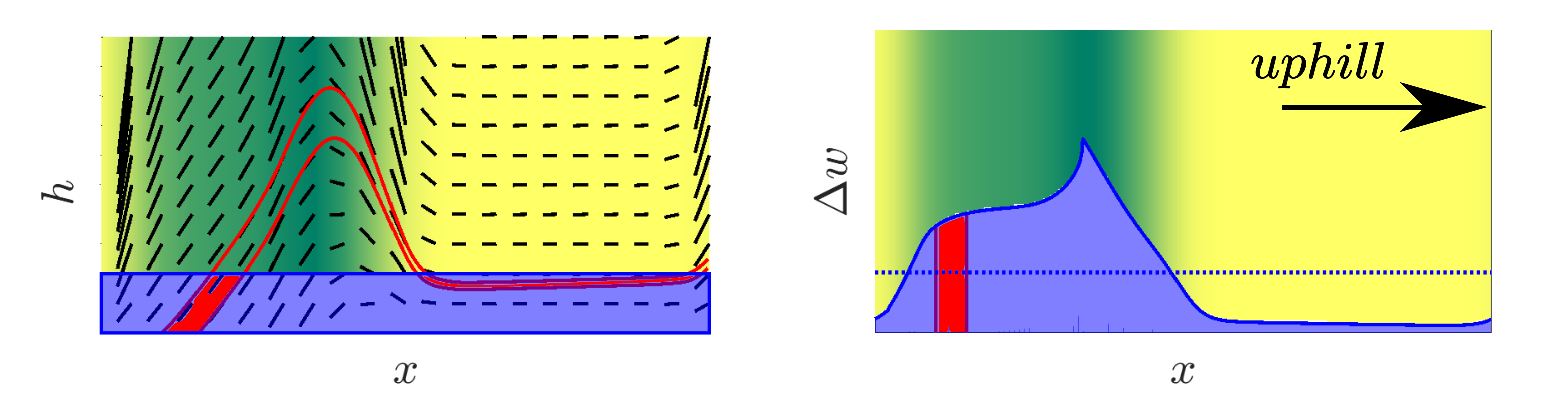}
    \caption{Illustration of the surface water transport and infiltration process modeled by the fast system~\eqref{eq:fast:nondim}.  The initial surface water, shaded blue and red in the left panel, is redistributed by the dynamics of the fast system to a soil water distribution indicated by the corresponding shaded regions in the right panel. The spatial biomass profile is indicated with high/low levels in green/yellow in the upper portions of both panels.  The direction field associated with the characteristic equation~\eqref{eq:charex} is indicated in black in the left panel, with two example curves in red. The red region between the two example curves, constrained to be below the initial height $h_0$, determines the soil water in red in the right panel. As indicated by the arrow in the right panel,  $x$ increases in the uphill direction.
    \label{fig:FastIntro}}
\end{figure}

The goal of this subsection is to obtain a closed form expression for $\widehat{w}(x)$, which is the amount of soil water at each location $x$, after the rain event.   This distribution, together with $\hat{b}(x)$ is then the initial condition for the slow subsystem \eqref{eq:slow:nondim} that applies during the ensuing dry spell.  Before going into the details of the calculation, we present a schematic summary of the result in Figure~\ref{fig:FastIntro}.  In particular, this figure illustrates the redistribution and infiltration processes that determine the amount of water from the rain event that infiltrates into a given location.      
The biomass level is indicated with high (low) levels shaded in green (yellow) in the upper part of both panels.  The left panel illustrates the initial surface water input by the blue (and red) shaded region below some initial surface water height $h_0$ in the positive $(x,h)$ quadrant.  This initial amount of water is redistributed by downhill surface transport and infiltration, resulting in the soil water distribution indicated by the blue (and red) shaded region in the right panel. Specifically, the amount of the initial block of surface water between the two red curves in the left panel, filled in red, infiltrates the soil at the locations where those curves reach $h=0$. This contribution to soil moisture is summarized by the corresponding red block in the right panel. Note that the added soil water distribution $\Delta w$, shown in the right panel, is concentrated where the biomass is located, reflecting the positive feedbacks of the system.

\begin{figure}
    \centering
    \includegraphics[width=0.85\textwidth]{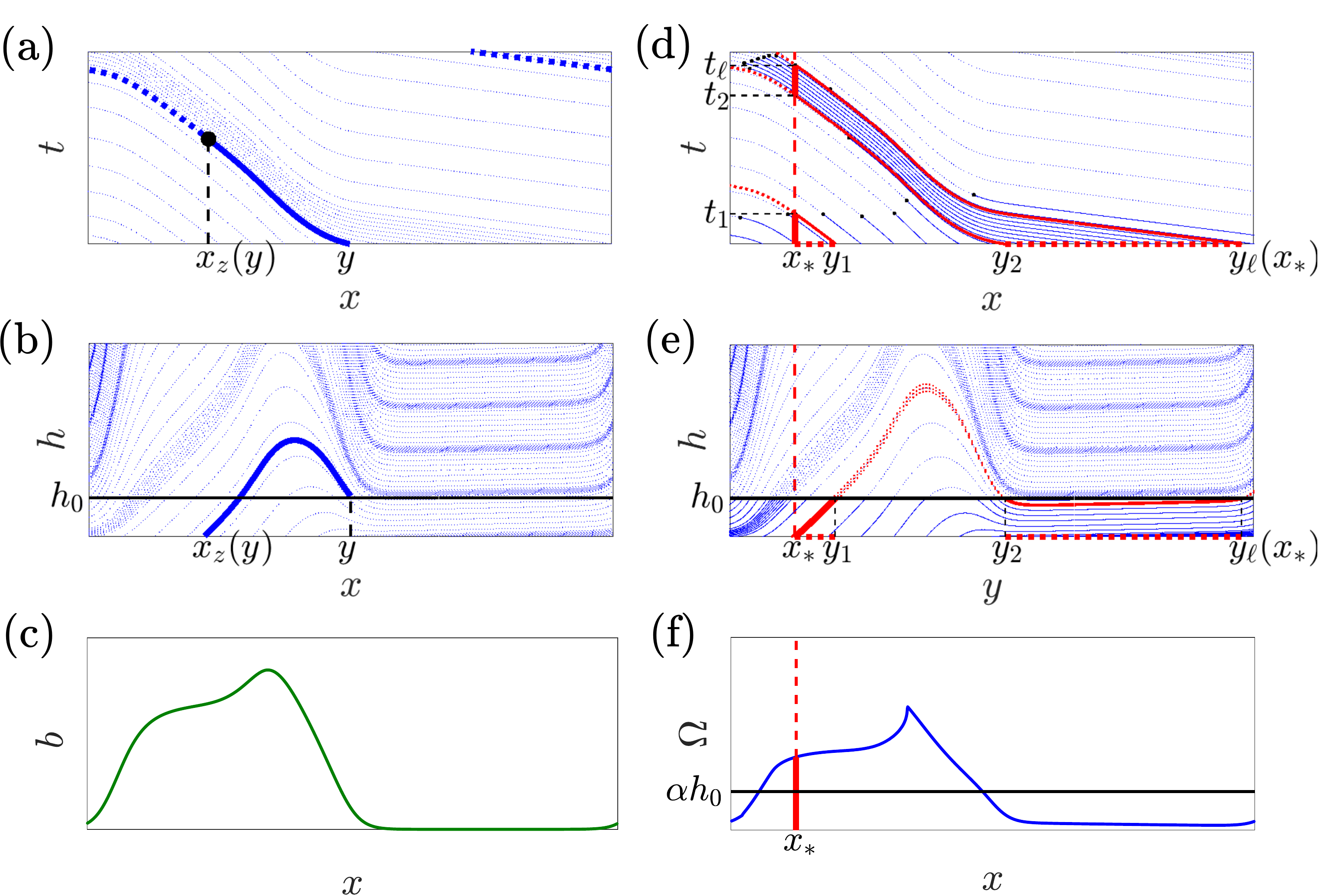}
    \caption{(a) The space time characteristic, defined by equation~\eqref{eq:txy}, with initial condition $h(y,0)=h_0$ is highlighted in bold.  Surface water $h$ reaches zero at $x_z(y)$ along the characteristic, which is indicated by a transition from solid to dotted line. (b) The corresponding picture in the $(x,h)$ plane shows how $h$ evolves along the characteristic, labeled by the location $y$ of the initial height $h_0$.  (c) The biomass profile is included to highlight it's role of slowing water flow and increasing infiltration. (d) The space time characteristics for which $h$ reaches zero at $x=x_*$ are highlighted in red.  The (vertical) time intervals  $(0,t_1)$ and $(t_2,t_\ell)$ at $x=x_*$ capture the times for which $h>0$ at $x=x_*$ and infiltration occurs.  (e) The corresponding picture in the $(y,h)$ plane indicates that the contributions to the soil water at $x=x_*$ come from locations with $h<h_0$ along the characteristic labeled by $y=y_\ell(x_*)$ which, by definition~\eqref{eq:hhat0}, reaches $h=0$ at $x_*$.  
    (f) The contribution from the rainstorm  to the soil water at $x=x_*$, determined by equation~\eqref{eq:Omegax0}, is highlighted in red. 
    }
    \label{fig:SolveFast}
\end{figure}

In this section, we describe how we obtain the red curves that define the redistribution process as the solution of system~\eqref{eq:fast:nondim} by a standard application of the method of characteristics (see, e.g.~\cite{evans2010pdes}). In particular, the solutions to the $h$ equation~\eqref{eq:fast:nondim:h} along so-called characteristic curves, defined below by~\eqref{eq:txy}, partition the initial water by where it ends up in the soil as illustrated in Figure~\ref{fig:FastIntro}. 
Figure \ref{fig:SolveFast} provides further details behind our approach for the biomass profile, ${\cal B}(X)\equiv Q\hat{b}(x)$, shown in panel (c). It fills in the steps to our geometric interpretation (Figure \ref{fig:FastIntro}) for the way water, initialized on the surface, gets redistributed into the soil. 
While the following discussion of our method assumes an infinite domain to make the  presentation of the underlying ideas more clear, our simulations incorporate periodic boundary conditions on a domain of length $L$. 

We employ the method of characteristics to solve the fast subsystem \eqref{eq:fast:nondim}, given a biomass distribution $\hat{b}(x)$, which determines the infiltration rate $\iota(x)$ and overland flow speed $\nu(x)$ via \eqref{eq:iotanu}. We assume that the initial surface water height is $h(x,0)=h_0$, which is set by the precipitation pulse, and that the initial soil water distribution is $w(x,0)=w_0(x)$. We begin with the $h$ equation \eqref{eq:fast:nondim:h}, which is decoupled from the $w$ equation~\eqref{eq:fast:nondim:w} thanks to our simplifying assumptions related to the infiltration function ${\mathcal{I}}$. We parameterize time $t(x;y)$ by spatial position $x$ along a characteristic that starts at $x=y$ at time $t=0$. This results in the following set of ODEs, one for each $\hat{h}(x;y)\equiv h(x,t(x;y))$, the height of the surface water along the characteristic starting at $y$:
\begin{equation}
\label{eq:charex}
    \frac{d}{dx}\Bigl(\nu(x)\hat{h}(x;y)\Bigr)=\iota(x)
    \Theta\Bigl(\hat{h}(x;y)\Bigr),\quad \hat{h}(y;y)=h_0,
\end{equation}
where the time to reach the position $x\le y$ along the characteristic starting at $y$ is given by
\begin{equation}
\label{eq:txy}
    t(x;y)=\int_x^y \frac{1}{\nu(s)}ds.
\end{equation}
Based on the functional forms, we can assume $\nu(x)$ and $\iota(x)$ are continuous and strictly positive. Therefore, by equation~\eqref{eq:charex}, the product $q(x;y)=\nu(x)\hat{h}(x;y)$, which is initially positive at $x=y$, decreases monotonically as $x$ decreases, i.e. in the downhill direction.  It can reach zero only when $\hat{h}(x;y)=0$, and for $x$ values below this point, $\hat{h}(x;y)$  remains zero, a consequence of the Heaviside function in equation~\eqref{eq:charex}. 
For the characteristic starting at $y$, we denote that point where the surface water reaches zero by $x_z(y)$. We can then integrate equation~\eqref{eq:charex} from the start of the characteristic $y$ to some point $x$ to get
\begin{equation}
\label{eq:hhat}
    \hat{h}(x;y)\equiv h(x,t(x,y)) =\begin{cases}
    \frac{1}{\nu(x)}\left(\nu(y)h_0-\int_{x}^{y}\iota(s)ds \right)& {\rm if}\ x_z(y)<x\le y\\
    0, &{\rm if}\ x\le x_z(y)
    \end{cases},
\end{equation}
Figure~\ref{fig:SolveFast}(a) illustrates a number of characteristics $t(x;y)$ satisfying \eqref{eq:charex}.  (Note that due to the periodic boundary conditions the characteristics wrap around the domain.)  
We indicate the point on the characteristic where $\hat{h}(x;y)$ first reaches zero,  $(x_z(y),t(x_z(y);y))$,  by a black circle and change to a dotted line where  $\hat{h}(x;y)=0$. 

With a solution for $h$ in hand, we now turn to the soil water equation~\eqref{eq:fast:nondim:w}. We can write a formal solution for $\widehat{w}(x)$ in terms of $h(x,t)$ as
\begin{equation}
    \label{eq:what}
    \widehat{w}(x)=w_0(x)+\underbrace{\alpha \iota(x)
    \int_0^\infty \Theta(h(x,t))dt}_{\equiv\Omega(x),\ {\rm added\ soil\ water}}.
\end{equation}
Here the integral over the Heaviside function  determines the total length of time surface water is infiltrating the soil after a rain pulse, on the fast timescale.  
Figure~\ref{fig:SolveFast}(d) indicates, with thick red vertical line segments, two intervals that comprise this time for the given point $x=x_*$.  In order to compute the integral in equation~\eqref{eq:what} given our solution $\hat{h}(x;y)$ in equation~\eqref{eq:hhat}, we make the change of variables from time $t$ to the starting position $y$ of the characteristic that reaches $x$ at time $t(x;y)$ via equation~\eqref{eq:txy}.   The added soil water is given by
\begin{equation}
\label{eq:Omegax}
    \Omega(x)=\alpha\iota(x)
    \int_{x}^{\infty}\frac{\Theta\Bigl(\hat{h}(x;y)\Bigr)}{\nu(y)} dy.
\end{equation}
The positive contributions to the integral in equation~\eqref{eq:Omegax} occur for values of $y$ where the surface water height $h$ is nonzero at $x$ along the characteristic starting at $y$. There are two such intervals in $y$ associated with $x=x_*$ for the example in Figure~\ref{fig:SolveFast}(d), which are indicated with thick dotted red lines along the spatial axis.  The characteristics highlighted in red in Figure~\ref{fig:SolveFast}(d), have surface water height that just reaches 0 at $x=x_*$ and mark the boundaries of the contributing intervals for both equations~\eqref{eq:what} and~\eqref{eq:Omegax}.  Figure~\ref{fig:SolveFast}(f) shows the resulting soil water distribution $\Omega(x)$ added by the impulse of rain.  The contribution at $x=x_*$ highlighted by a thick solid red line is proportional to the time during which infiltration occurs.

In order to develop a geometric interpretation of equation~\eqref{eq:Omegax}, we return to our set of ODEs for $\hat{h}(x;y)$, with each labeled by $y$ and given in equation~\eqref{eq:charex}.   Notice that it is possible for multiple characteristics to first reach zero surface water height at the same location, that is $x_z(y_1)=x_z(y_2)$ for $y_1\neq y_2$.  
Because we assume the surface water is initially uniformly distributed across the entire domain, there is at least one characteristic that first reaches zero at any point $x$ on the domain. We define $y_\ell(x)$ to be the largest $y$ such that $x_z(y)=x$. Now, integrating equation~\eqref{eq:charex}, along this $y_\ell(x)$ characteristic, from $x$ where the surface water reaches zero up to some point $y<y_\ell(x)$ gives
\begin{equation}
\label{eq:hhat0}
\hat{h}(y;y_\ell(x))=\frac{1}{\nu(y)}\int_x^y\iota(s) ds.    
\end{equation}
The integral in equation~\eqref{eq:hhat0} represents the height of the surface water at a point $y$ along a characteristic defined by equations~\eqref{eq:charex} and~\eqref{eq:txy} that starts at $x$ with $h=0$ and follows it backward in time.  We use the same notation $\hat{h}$ as in equation~\eqref{eq:hhat} here because the result of the integral in~\eqref{eq:hhat0} is equivalent to the surface water height at the point $y$ along the same characteristic curve, but starting at $y_\ell(x)$ with $h=h_0$ and integrating forwards in time. 

Noting that a positive multiplicative factor $\nu(x)/\nu(y)$ can be inserted 
in the argument of the Heaviside step function of equation~\eqref{eq:charex}, we can  make use of equation~\eqref{eq:hhat} together with~equation~\eqref{eq:hhat0} to re-express the added soil water $\Omega(x)$ as
\begin{equation}
\label{eq:Omegax0}
      \Omega(x)=\alpha\iota(x)
\int_{x}^{y_\ell(x)}\frac{\Theta\Bigl(h_0-\hat{h}(y;y_\ell(x))\Bigr)}{\nu(y)} dy,
\end{equation}
We can truncate the upper bound of the integral in going from equation~\eqref{eq:Omegax}~to~\eqref{eq:Omegax0} because we have defined $y_\ell(x)$ such that $\hat{h}(y;y_\ell(x))>h_0$ for all $y>y_\ell(x)$.  

Equation~\eqref{eq:Omegax0} affords a geometric picture for the water going from the surface into the soil.  We imagine beginning with a ``block" of water on the surface as a result of the rainstorm, which is represented by the region between the thick solid black line and the $y$-axis in the $(y,h)$-plane in Figure \ref{fig:SolveFast}(e).  The fast system acts to redistribute this initial block of water into the soil via surface transport and infiltration.  We can think of the characteristics defined by equation~\eqref{eq:hhat0} that start at $h=0$ and move backward in time as a partitioning of this block in the $(y,h)$-plane. The amount of the initial block of water that appears along a characteristic, starting at a location $x$ with $h=0$, is how much water infiltrates into that location.  Notice that the contribution to $\Omega(x)$ along a characteristic is limited by the Heaviside function to intervals where $h<h_0$, e.g. within the initial block of water. Typical characteristics are indicated in Figure~\ref{fig:SolveFast}(e) by blue lines, with the intervals above $h_0$ dotted.  The characteristic that starts at $x=x_*$ and determines the amount of added soil water at that location is highlighted in red.  Notice that there are two intervals of this characteristic below $h_0$. The endpoints of these segments correspond exactly to the endpoints of the intervals of integration $(x_*,y_1)$ and $\bigl(y_2,y_\ell(x_*)\bigr)$ shown in Figure~\ref{fig:SolveFast}(d).
Indeed, all the characteristics that start with $h=h_0$ at $t=0$ and reach zero at $x=x_*$ in the $(x,t)$-plane of Figure~\ref{fig:SolveFast}(d) map onto segments of the characteristic that starts with $h=0$ at $x=x_*$ in the $(y,h)$-plane  of Figure~\ref{fig:SolveFast}(e).    

The numerical simulations reported on in Sections~\ref{sec:per} and~\ref{sec:stoch} are carried out using Matlab's ODE suite~\cite{shampine1997matlab}. The contribution to the soil water from each rain pulse is computed by numerical integration of equations~\eqref{eq:hhat0}-\eqref{eq:Omegax0} via the trapezoidal rule. A centered finite-difference scheme is used to evolve the slow system~\eqref{eq:slow:nondim} in between the rain pulses.

\section{Periodic Rainfall}
\label{sec:per}

We begin exploration of the pulsed-precipitation model
by first considering an evenly-spaced sequence of identical rain events within each rainy season, which is the periodic case shown in Figure~\ref{fig:rainfall}(f).
We find that the  regularity of this artificial rainfall pattern leads to a spatial resonance phenomenon that controls the preferred spacing of the vegetation bands. Specifically, the spacing is determined by the distance surface water can travel in the time it takes for the precipitation pulse to fully infiltrate into the soil. We show this explicitly in Section~\ref{sec:per:linstab}, through a linear stability analysis of the uniform vegetation state to spatially periodic perturbations proportional to $e^{ikx}$. This analysis reveals  a sequence of resonance tongues in a $({\cal MAP},k)$-parameter plane, where ${\cal MAP}$ denotes mean annual precipitation. We then use this insight to understand the preferred spacing and travel direction of the fully nonlinear bands, which is obtained through numerical simulations of the model 
in Section~\ref{sec:per:nonlin}. 

Throughout this section we use a precipitation model consisting of $N_s=2$ identical rainy seasons, six months apart, so the periodicity is $T_p=(365/2)\ days$. Each rainy season lasts for a time $T_r=(365/12)\ days\equiv 1\ month$, and it consists of $N_p$ equally spaced rain pulses that deliver, instantaneously, a column of water of height $H_0$. The mean annual precipitation is then ${\cal MAP}=N_s N_p H_0$.
We denote dimensionless time units by $\tau$, with the six month period of the seasonal forcing given by $\tau_p=1.825$, and the month-long rainy season lasting a time $\tau_r=\tau_p/6$. The precipitation pulses, in dimensionless units, are denoted by $h_0$.

\subsection{Linear stability of spatially uniform, temporally-periodic solutions}
\label{sec:per:linstab}

In this subsection, we describe results of linear stability computations for uniform vegetation to heterogeneous perturbations, proportional to $e^{ikx}$. The uniform state has the same half-year periodicity as the rainfall pattern and is determined as a fixed point of an appropriate stroboscopic map. Its linear stability properties are determined by a computation of Floquet multipliers as a function of $k$.

\paragraph{Loss of Stability of Bare Soil Solution}
We find that the uniform vegetation state arises from a transcritical bifurcation of the zero-biomass desert state at ${\cal MAP}={\cal MAP}_c\equiv LM/C\Gamma$. In particular, as shown in  Appendix~\ref{app:transcrit}, this threshold is independent of the details of the rainfall model. 
In fact, if we were to replace our fast-slow system for the uniform solutions by a pair of ordinary differential equations for slow variables $(W(T),B(T))$, with a constant precipitation rate $P_0$, then we would obtain the same instability boundary. Specifically, we find that there is a transcritical bifurcation, which produces the uniform vegetation solution, when $P_0=P_c=LM/C\Gamma={\cal MAP}_c$ for 
\begin{eqnarray*}
\dot{W}&=&P_0-(L+\Gamma B)W\\
\dot{B}&=&C\Bigl(1-\frac{B}{K_B}\Bigr)\Gamma B W-MB.
\end{eqnarray*}
This transcritical bifurcation marks the stability boundary for the zero-biomass desert state; it's unstable for ${\cal MAP}>{\cal MAP}_c.$
Numerical simulations using parameters given in Table~\ref{tab:dim} indicate that patterns may stably co-exist with desert well below
${\cal MAP}_c\approx 11\, cm$. Our numerical investigations of pattern collapse  reported in Section~\ref{sec:stoch:col} are carried out in such a stable co-existence regime, using a stochastic rainfall model with an average ${\cal MAP}$ of $8\ cm$.

\paragraph{Pattern-Forming Instability of the Uniform Vegetation Solution} In contrast to the desert state, we find that the stability region for the uniform vegetation state, to heterogeneous perturbations, depends on details of the rainfall model. 
Here we describe our linear stability calculations and summarize some of the key findings related to pattern-forming instabilities of the uniform vegetation state. 

Let $(w_0,b_0)$ denote soil water and biomass levels of the $\tau_p$-periodic uniform vegetation state at the start of the rainy season.
This state exists, with $b_0>0$, for ${\cal MAP}>{\cal MAP}_c$. We evolve it, together with a small spatially periodic perturbation $(\Delta w_{k,0},\Delta b_{k,0})e^{ikx}$, over one cycle of the  periodic rainfall model. The (linearized) Poincar\'e return map, 
\begin{equation}
\label{eq:poincaremap}
   {\cal P}_{\tau_p,k}(w_0,b_0,\Delta w_{k,0},\Delta b_{k,0})=(w_0,b_0,\Delta w_{k,1},\Delta b_{k,1}), 
\end{equation}
has fixed point $(w_0,b_0,0,0)$. This map also determines the linear evolution of the  perturbation $(\Delta w_{k,0},\Delta b_{k,0})$, over the period $\tau_p$, to its updated value $(\Delta w_{k,1},\Delta b_{k,1})$. We quantify this change by a pair of Floquet multipliers, which are the eigenvalues of the linearized Poincar\'e return map restricted to the perturbations. We denote this two-dimensional linear map by ${\cal L}_{k}$, i.e.
 \begin{equation}
 \label{eq:L}
     \begin{pmatrix}
      \Delta w_{k,1}\\
      \Delta b_{k,1}
     \end{pmatrix}={\cal L}_{k}
      \begin{pmatrix}
      \Delta w_{k,0}\\
      \Delta b_{k,0}
     \end{pmatrix}.
 \end{equation}

The return map, ${\cal P}_{\tau_p,k}$, is a composition of maps associated with the fast and slow subsystems. The fast subsystem distributes the water in the soil after a rain pulse of strength $h_0$. The associated map, denoted $\psi_{h_0,k}$, takes the form
\begin{equation}
\label{eq:psihk}
    \psi_{h_0,k}(w,b,\Delta w_k,\Delta b_k)=(w+\alpha h_0,b,\Delta w_k+\Delta \Omega_{h_0,k},\Delta b_k).
\end{equation}
Note that the initial biomass $b+\Delta b_ke^{ikx}$ is frozen for the fast system and hence is unchanged.  The soil moisture gains a uniform contribution $\alpha h_0$ and, for the linearized problem, a nonuniform contribution $\Delta \Omega_{h_0,k}\ e^{ikx}$. We determine the latter by linearizing \eqref{eq:Omegax0} about $\Delta b_k=0$, and setting $y_\ell(x)=x+\ell_0+\delta\ell_k(x)$, where
\begin{equation}
    \label{eq:ell0}
\ell_0=\frac{\nu(b) }{\iota(b)}\ h_0
\end{equation}
is the distance surface water of height $h_0$ travels, over uniform biomass at level $b$, before completely infiltrating into the soil. The change in this travel distance due to the biomass perturbation $\Delta b_ke^{ikx}$, denoted $\delta\ell_k(x)$, is  determined below. First, to linear order in $\Delta b_k$ and $\delta\ell_k(x)$, we find
\begin{eqnarray}
\label{eq:omegax}
\Omega(x)
&=&
\alpha \iota(b+\Delta b_k e^{ikx})\int_x^{y_\ell(x)}\frac{\Theta(h_0-\widehat{h}(y;y_\ell(x)))}{\nu(b+\Delta b_ke^{iky})}\ dy\nonumber\\
&=&\alpha\Bigl(\iota(b)+\frac{d\iota}{db}\Delta b_ke^{ikx}\Bigr)\int_x^{x+\ell_0+
\delta\ell_k(x)}\Bigl(\frac{1}{\nu(b)}-\frac{1}{\nu(b)^2}\frac{d\nu}{db}\Delta b_ke^{iky}\Bigr) dy+\cdots\nonumber\\
&=& \alpha h_0+\underbrace{\alpha h_0\Bigl[\frac{\delta\ell_k(x)}{\ell_0}+\Bigl(\frac{1}{\iota(b)}\frac{d\iota}{db}+ \frac{i(e^{ik\ell_0}-1)}{\nu(b)k\ell_0}\frac{d\nu}{db}\Bigr)\Delta b_ke^{ikx}\Bigr]}_{=\Delta\Omega_{h_0,k}\ e^{ikx}}+\cdots,
\end{eqnarray}
where the ellipsis refers to higher order terms in $\Delta b_k$ and $\delta\ell_k$. To find the slight adjustment, $\delta\ell_k(x)$, to the total travel distance to complete infiltration at location $x$, which is specifically due to the biomass perturbation $\Delta b_ke^{ikx}$, we solve $\widehat{h}(y_\ell(x);y_\ell(x))=h_0$ using \eqref{eq:hhat0}. We find, to linear order in $\Delta b_k$,
\begin{equation}
\label{eq:deltalx}
    \frac{\delta\ell_k(x)}{\ell_0}=\Bigl[\frac{1}{\nu(b)}\frac{d\nu}{db}e^{ik\ell_0}+\frac{i}{\iota(b)k\ell_0}\frac{d\iota}{db}\Bigl(e^{ik\ell_0}-1\Bigl)\Bigr]\Delta b_ke^{ikx}.
\end{equation}
Combining \eqref{eq:omegax}-\eqref{eq:deltalx}, we obtain, for $k\ne 0$,
\begin{equation}
\label{eq:domega}
    \Delta\Omega_{h_0,k}=\alpha h_0 \left[ \frac{1}{\iota(b)}\frac{d \iota}{d b}
    + \frac{1}{\nu(b)}\frac{d \nu}{d b} e^{ik\ell_0}
    +\frac{i}{k \ell_0}\left(\frac{1}{\iota(b)}\frac{d \iota}{d b} + \frac{1}{\nu(b)}\frac{d \nu}{d b}\right)\left(e^{ik\ell_0}-1\right)
    \right] \Delta b_k.
 \end{equation}
Here
$\ell_0$ is given by \eqref{eq:ell0}, and
\begin{equation*}
    \frac{1}{\iota(b)}\frac{d \iota}{d b}=\frac{(1-f)}{(b+f)(b+1)},\quad       \frac{1}{\nu(b)}\frac{d \nu}{d b}=-\Bigl(\frac{\eta}{1+\eta b}\Bigr)
\end{equation*}
follow from \eqref{eq:iotanu}.
(Note that for $k=0$, it can be shown that  
$\Omega_{h_0,0}=0$.)

The flow map that applies between pulses is derived from the slow system \eqref{eq:slow:nondim}. It consists of the nonlinear equations satisfied by the uniform vegetation state $(w,b)$,  together with the linear  equations in the perturbations $(\Delta w_k,\Delta b_k)$. We denote this map, which flows from an initial condition $(w_0,b_0,\Delta w_{k,0},\Delta b_{k,0})$ for a time $\tau$, by
\begin{equation}
\label{eq:psitauk}
    \varphi_{\tau,k}(w_0,b_0,\Delta w_{k,0},\Delta b_{k,0})=(w(\tau),b(\tau),\Delta w_k(\tau),\Delta b_k(\tau)).
\end{equation}
Here
$w(\tau),b(\tau),\Delta w_k(\tau),\Delta b_k(\tau)$ satisfy
\begin{subequations}
	\label{eq:slow:odes:linear:k}
\begin{align}
\frac{d w}{d \tau}&= -(\sigma+\gamma b) w\label{eq:slow:odes:linear:k:w}\\
\frac{d b}{d \tau} &= \left(1- \frac{b}{\kappa}\right)w b - b. \label{eq:slow:odes:linear:k:b}\\
\frac{d \Delta w_k}{d \tau}&=  -(\delta_w k^2 + \sigma+\gamma b)\Delta w_k-\gamma w\Delta b_k \label{eq:slow:odes:linear:k:dw}\\
\frac{d \Delta b_k}{d \tau} &= \left(1- \frac{b}{\kappa}\right) b\Delta w_k+ \left(-\delta_b k^2 + w -\frac{2b}{\kappa}w-1\right)\Delta b_k. \label{eq:slow:odes:linear:k:db}
\end{align} 
\end{subequations}

We can now construct the Poincar\'e return map \eqref{eq:poincaremap} as 
\begin{equation}
    \label{eq:poincare2}
    {\cal P}_{\tau_p,k}=\varphi_{\tau_d,k}\circ
    \underbrace{(\varphi_{\Delta \tau,k}\circ\psi_{h_0,k})\circ\ldots\circ(\varphi_{\Delta \tau,k}\circ\psi_{h_0,k}) }_{N_p \ {\rm times}}.
\end{equation}
Here $\Delta\tau=\tau_r/N_p$, which is the time between rain pulses during the rainy season, while $\tau_d=\tau_p-\tau_r$ is the length of the dry season. Restricting the map to the two perturbation components determines ${\cal L}_k$ in \eqref{eq:L}.
The eigenvalues, $\lambda_k$, of ${\cal L}_k$ are the (complex) Floquet multipliers. (Note that it follows from \eqref{eq:domega} that the map itself has complex entries.)  If the modulus of either eigenvalue $\lambda_k$ exceeds one, then the uniform state is unstable to pattern-forming perturbations of wavenumber $k$. 

\begin{figure}
    \centering
    \includegraphics[width=\textwidth]{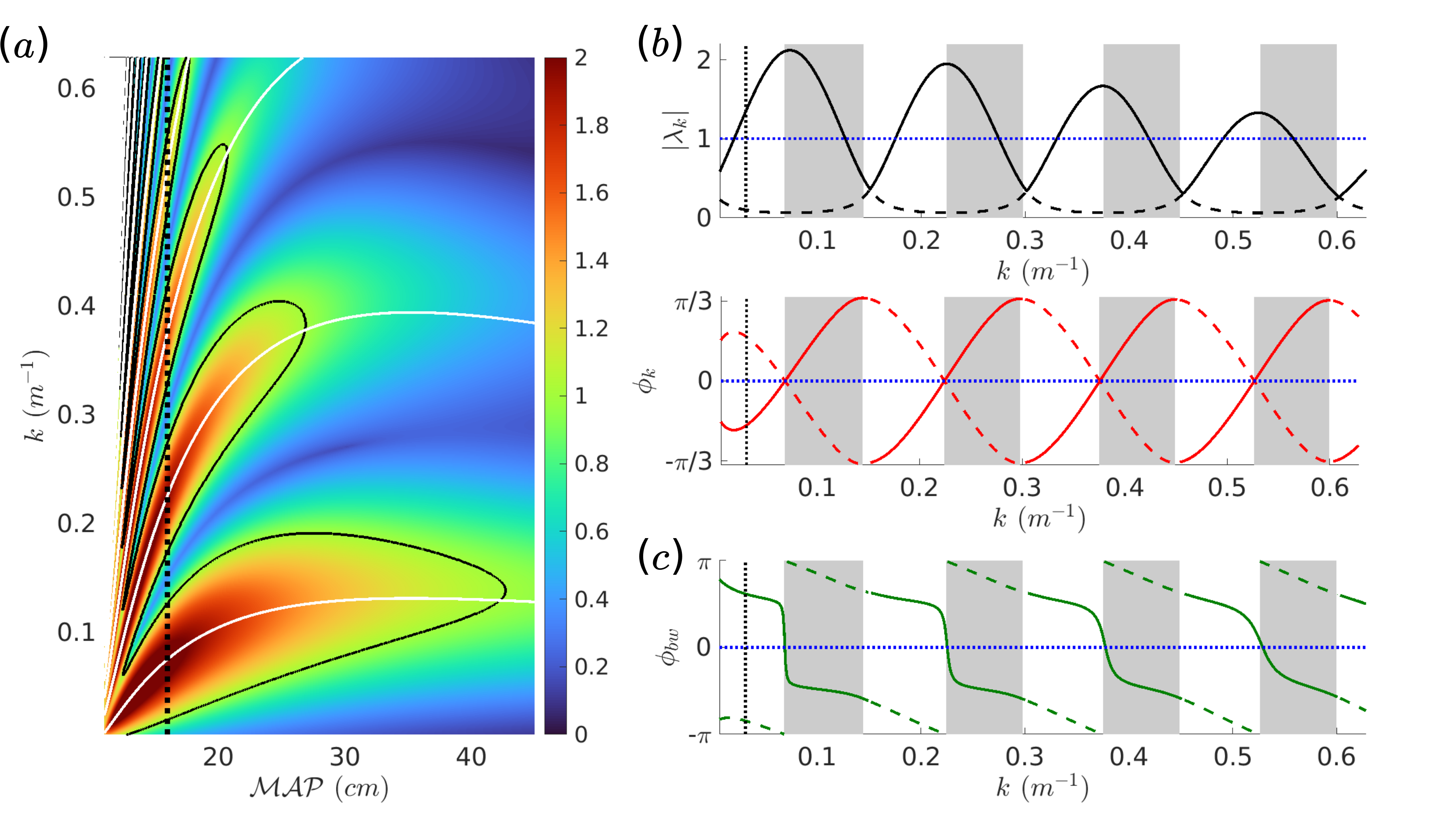}
    \caption{
    (a) Magnitude of leading eigenvalue of $\mathcal{L}_k$, defined by equation~\eqref{eq:L}, as a function of mean annual precipitation (${\cal MAP}$) and perturbation wavenumber $k$.  Solid black lines are linear stability boundaries and solid white lines indicate predicted resonances between the pattern wavelength and characteristic surface flow distance. 
    (b) The magnitude and phase of the two eigenvalues $\lambda_k$ along the dotted black line at ${\cal MAP}=16$ $cm$ in panel (a).  The solid lines correspond to the larger magnitude eigenvalue and the dashed correspond to the smaller one. Shaded regions yield downhill migration of perturbations, inferred from the sign of $\phi_k$ (c) The phase of the biomass perturbation relative to the soil water perturbation associated with the eigenvectors for eigenvalues shown in (b).   The wavenumber 
    corresponding to a wavelength of $200\,m$ is marked by black vertical dotted lines in right panels; compare to nonlinear pattern with same wavenumber in Figure~\ref{fig:nonlinPer}.   
    }
    \label{fig:linstab0Rk}
\end{figure}

Figure~\ref{fig:linstab0Rk} presents, in dimensioned quantities, an example of typical linear stability results for the parameters of Table~\ref{tab:dim}. To obtain these results, we first numerically compute the uniform vegetation fixed point $(w_0,b_0,0,0)$ of the Poincar\'e return map~\eqref{eq:poincaremap}, and then numerically compute its linear stability via the eigenvalues $\lambda_k$ of the pertubation map~\eqref{eq:L}. Figure~\ref{fig:linstab0Rk}(a) shows a heat map of the largest $|\lambda_k|$ as a function of both mean annual precipitation ${\cal MAP}$ and perturbing wavenumber $k$ in the case of a fixed number of rainstorms per season $N_p=8$. (For $N_p=8$, storm strengths range between $H_0\approx 0.68\ cm$ and $H_0\approx 2.7\ cm$ for Figure~\ref{fig:linstab0Rk}(a).)
We find that as ${\cal MAP}$ decreases from a high value of ${\cal MAP}=45\ cm$, by continuously decreasing $H_0$, the uniform state loses stability at ${\cal MAP}\approx 42.8\, cm$ to perturbations of  wavenumber $k\approx 0.141\, m^{-1}$ (wavelength $L\approx 45\, m$). 
This figure, which shows the instability boundary associated with $|\lambda_k|=1$ as a black curve, captures a structure in the form of ``resonance tongues". Specifically, we find that the instability regions  straddle predictions based on overland water flow distances $L_n=2\ell_0/(n+1)$, $n=0,1,2,\ldots$ (solid white lines), where $\ell_0$ is given by \eqref{eq:ell0}. (For this we evaluate  $\nu(b)$ and $\iota(b)$ in $\ell_0$ using the fixed point value $b_0$ associated with the return map~\eqref{eq:poincaremap}, i.e. its level at the start of  the rainy season.) We find that the most unstable perturbation has a wavelength $L_0$ that is well-approximated by twice the distance water flows on the surface before getting infiltrated, i.e.  to the perturbation wavenumber $k=\pi/\ell_0$.   Thus, at the linear level, the ``preferred" pattern wavelength is one for which the newly-forming vegetation band harvests water from the location of the newly-forming bare soil region.  This instability appears as part of a series of increasingly weaker and narrower instability regions, which correspond to the surface water from a newly-forming bare soil region traveling $n$ wavelengths before reaching a newly-forming vegetation band. Figure \ref{fig:linstab0Rk}(b) shows a plot of $|\lambda_k|$  for ${\cal MAP}=16\ cm$, indicated by a black dotted line in panel (a), which slices through four of the instability tongues of Figure \ref{fig:linstab0Rk}(a). 

In order to determine  linear predictions for vegetation band migration speed, we extract the phases $\phi_k$ of the eigenvalues $\lambda_k=\lvert \lambda_k \rvert e^{i\phi_k}$ of $\mathcal{L}_k$.
The second panel of Figure~\ref{fig:linstab0Rk}(b) shows a plot of  $\phi_k$, for both eigenvalues $\lambda_k$ for ${\cal MAP}=16\, cm$, and indicates that, for all four instability intervals, the phase switches sign  near each successive peak of $|\lambda_k|$, and that the phase is confined to an interval around $\phi_k=0$, here approximately $[-\pi/3,\pi/3]$. 
If the phase is negative for the unstable Floquet multiplier, then that indicates a phase advance of the pattern during each seasonal cycle, and thus corresponds to uphill migration of the vegetation bands. The opposite holds for a  positive phase, shaded gray in panels (b) and (c), which  indicates a downhill migration of the vegetation bands.
Figure~\ref{fig:linstab0Rk}(c) shows the  phase shift between the components $\Delta w_k$ and $\Delta b_k$ of the eigenvectors of ${\cal L}_k$, e.g. the eigenvector can be written $(\Delta w_k,\Delta b_k)=(1,{\cal R}e^{i\phi_{bw}})$, where ${\cal R}>0$ is real.
We see that the sign of the phases $\phi_k$ in Figure~\ref{fig:linstab0Rk}(b) are
(typically) opposite to the phase denoted $\phi_{bw}$
in Figure~\ref{fig:linstab0Rk}(c). 
This observation lends itself to a simple interpretation. Specifically, if $\phi_{bw}>0$, then the water peak is uphill  from the biomass peak and we might expect  uphill migration of the bands, i.e. $\phi_k<0$. Similarly, we expect $\phi_k>0$ (downhill migration) if $\phi_{bw}<0$, in which case the water peak is shifted downhill from the biomass one.

We note that the  prediction, of the linearized problem, that patterns might travel downhill is not consistent with observational studies, which report only upslope colonization. Moreover, the observed migration speeds for the bands are slow; for instance, order of magnitude, a band might take a century to migrate uphill by one wavelength~\cite{deblauwe2012determinants,gowda2018signatures}, which would correspond to a (negative) phase shift of $\sim \pi/100$ every seasonal cycle. In the subsequent sections we explore how well these linear findings hold up for the nonlinear problem under periodic and stochastic rain inputs.

\subsection{Nonlinear Patterns}
\label{sec:per:nonlin}

Numerical simulations indicate that, while the linear theory captures the behavior of small amplitude patterns near onset of the lowest order resonance tongue shown in Figure~\ref{fig:linstab0Rk}, the nonlinear patterns selected at lower ${\cal MAP}$ values exhibit very different dynamics from the linear predictions.  Even still, the importance of the characteristic distance surface water travels before infiltrating into the soil seems to carry over into the nonlinear regime, as we now demonstrate.

\begin{figure}
    \centering
    \includegraphics[width=\textwidth]{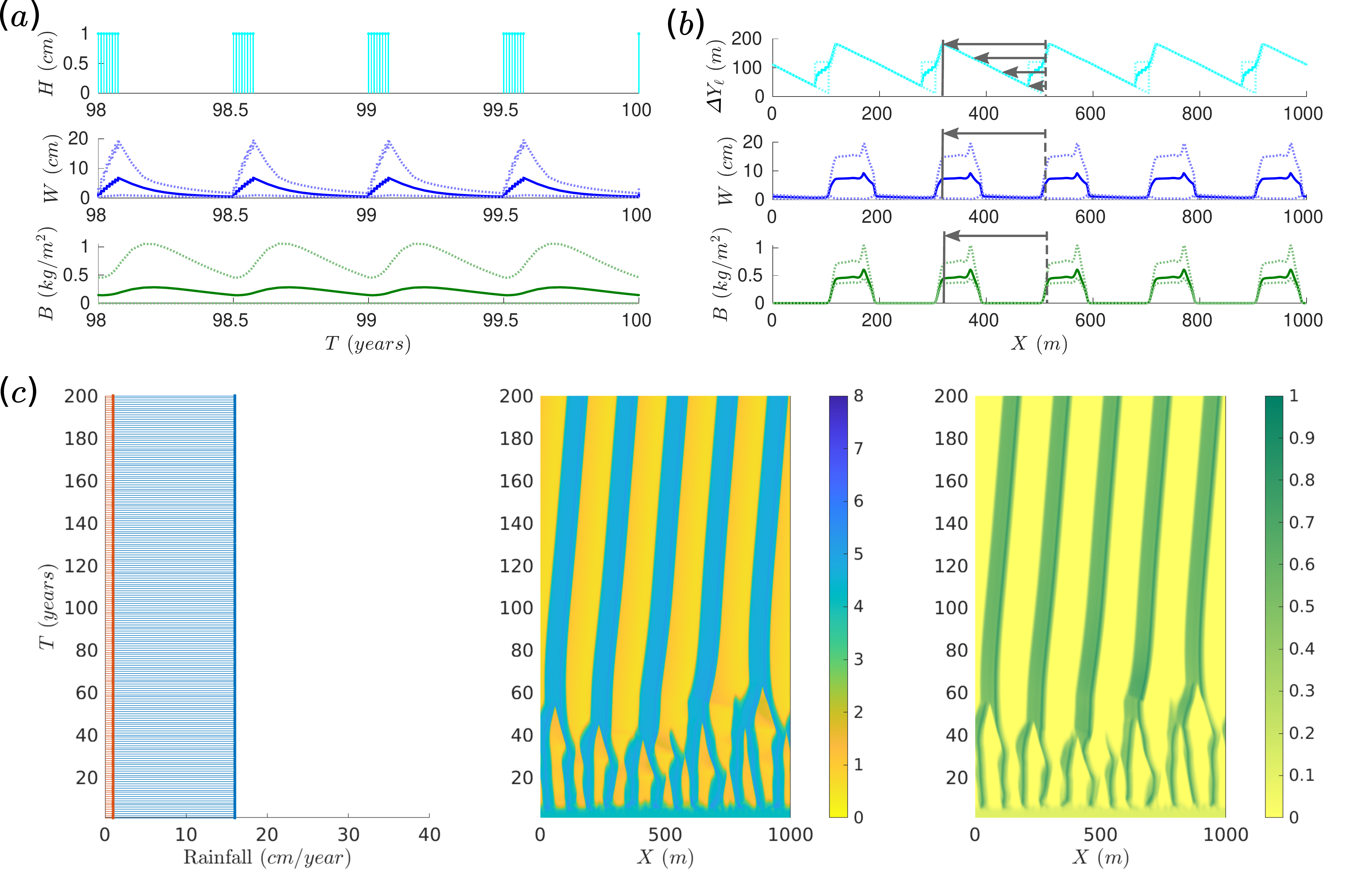}
    \caption{(a) Time series showing, from top to bottom, storm depth $H$ of rain pulses, domain-averaged soil water $W$ and biomass $B$ during last two years of a 100 year simulation under periodic rainfall. Solid lines indicate spatially averaged fields while dashed lines indicate instantaneous min/max. (b) Spatial distribution, from top to bottom, of dimensioned maximum distance $\Delta Y_\ell(X)$ traveled by water before infiltrating, soil water $W$ and biomass $B$ during the last year of the simulation. Solid lines indicate averages over rainstorms for $\Delta Y_\ell(X)$ and annual averages for $W$ and $B$.  Pointwise min/max values are shown by dashed lines. Horizontal gray arrows have length equal to $\Delta Y_\ell(X)$, and indicate farthest distance traveled by surface water infiltrating into locations at the arrow tips.  (c) From left to right, time series of annual rainfall totals in blue with $1\, cm$ contribution from each rainstorm highlighted in orange, spacetime plots of annually averaged soil water in units of $cm$ and biomass in units of $kg/m^2$. Parameters: Periodic rainfall with ${\cal MAP}=16\, cm$, storm depth $H_0=1\, cm$ and rainy season duration $T_r= 1\, month$ on a $L=1000\, m$ domain and initialized with 1\% random noise on top of the spatially uniform solution.}     \label{fig:nonlinPer}
\end{figure}

For Figure~\ref{fig:nonlinPer}, we consider periodic rainfall with ${\cal MAP}=16\, cm$ and storm depth of $H_0=1\, cm$ on a $1\, km$ domain.  Simulations initialized with the $0.1\%$ random noise on top of the uniform vegetation state typically settle into a ``traveling-wave'' state with five bands on the domain. Here we use ``traveling wave'' (in quotes) to indicate that the state undergoes a spatial translation under the nonlinear map associated with evolving the system by one rainy and subsequent dry season.  The five-band state (wavelength of $200\, m$, wavenumber of $\sim 0.0314\, m^{-1}$) has an uphill migration speed of  approximately $69\, cm/year$, corresponding to translation by one wavelength every $\sim 290$ years; this wavenumber is also indicated by a vertical line  in the linear results of Figure~\ref{fig:linstab0Rk}(b). Panel (a) provides timeseries data of spatially averaged quantities over the last two years of the 100-year simulation, panel (b) shows spatial profiles derived from the last year of the simulation and panel (c) shows annually-averaged spacetime plots.  We note that $\Delta Y_\ell(X)$ in panel (b) is the farthest (dimensioned) distance traveled by water on the surface during a rainstorm before infiltrating at a point $X$.  The dimensionless version of this quantity is $\Delta y_\ell(x)\equiv y_\ell(x)-x$, where $y_\ell$ is defined in Section~\ref{sec:pulsedprecip:solve} and sets the upper bound of integration for computing the soil water kick $\Omega$ in Equation~\eqref{eq:Omegax0}. The linear stability analysis of spatially uniform states in Section~\ref{sec:per:linstab} indicates that $\Delta y_\ell(x)=\ell_0$, a constant value in this case, plays a key role in controlling the wavelength of the patterns.  For the fully nonlinear patterns shown in Figure~\ref{fig:nonlinPer}, we see a ``sawtooth'' structure in the plot of the average $\Delta Y_\ell(x)$. Gray horizontal arrows indicate the farthest average distance traveled by water that infiltrates into locations between the peak value of $\Delta Y_\ell(X)=180\,m$ at $X=321\,m$ and the minimum value of $\Delta Y_\ell(X)=36\,m$ at $X=478\,m$. Water initialized at $X=514\,m$, within the trailing edge of a vegetation band, travels through nearly the entire vegetation band downhill before fully infiltrating.  
This  indicates that nearly the entire 86 $m$ width of the vegetation bands are harvesting water from bare soil regions uphill of them, and that none of the water travels across a band into the bare soil region downhill of it.

\begin{figure}
    \centering
    \includegraphics[width=\textwidth]{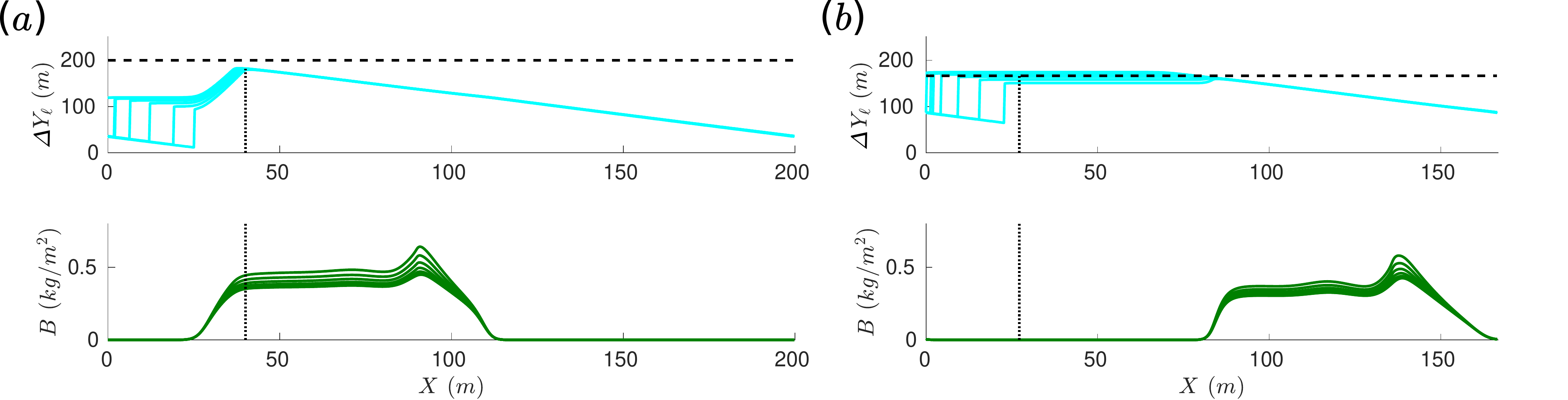}
    \caption{Spatial distribution of maximum distance  $\Delta Y_\ell(X)$ traveled by water before infiltrating and biomass profile associated with each of the eight rainstorms during one season with (a) $L=200\, m$ with average \textit{uphill} migration speed of $69\, cm/year$ and (b)  $L\approx 167\, m$ with average \textit{downhill} migration $179\, cm/year$. The location of the peak of $\Delta Y_\ell(X)$ averaged over the eight storms is marked by a vertical dotted line, while the domain size is marked by a horizontal dashed line. Parameters: Periodic rainfall with ${\cal MAP}=16\, cm$, storm depth $H_0=1\, cm$ and rainy season duration $T_r=1 \, month$. }
    \label{fig:persurflow}
\end{figure}

Using the same parameters, we also observe a state with 6 bands that migrate \textit{downhill} at an average rate of about 179 $cm/year$, or one $L\approx 167\, m$ wavelength every 93 $years$. Figure~\ref{fig:persurflow} zooms in on a single band of these 5-band and 6-band periodic patterns. It shows, from bottom to top, the spatial profiles of the biomass $B$
and the maximum distance $\Delta Y_\ell$ water travels before infiltrating into the soil, for each of the eight one-centimeter rainstorms of a rainy season.  The profiles
are aligned so that $X=0$ corresponds to the farthest downhill that water initialized within the vegetation band reaches during any of the rainstorms. With this choice, if there is a bare soil region uphill of the vegetation band, then we know that it did not collect water from a vegetated region during the rainy season.   

The uphill-migrating case with $L=200\, m$ in panel (a) is more ``optimal" in the sense that the water from the bare region uphill of the band is deposited completely within the band, as evidenced by the peak value of $\Delta Y_\ell$ being less than the domain length. Moreover, the peak is located within the vegetation band.  By contrast, the profile of $\Delta Y_\ell$ in the downhill migrating case with $L\approx 167\,m$, shown in panel (b), indicates that water is traveling
all the way across the band, and continuing into the downhill bare soil region. The peak values for $\Delta Y_\ell$ in this case, which is about $174\, m$, slightly exceeds the domain length.

\section{Stochastic Rainfall}
\label{sec:stoch}

In this section, we explore some of the striking differences in behavior of the pulsed precipitation model under stochastic rainfall, compared to the idealized periodic rainfall  results of Section~\ref{sec:per}. 
Numerical simulations indicate that banded patterns still form under stochastic rainfall, with characteristics  consistent with observational data for band spacing and migration speed.  Moreover, some of the more complex, and perhaps worrisome, spatiotemporal behaviors produced by the model with idealized periodic rainfall vanish once  stochasticity in rainfall is introduced.  We also find that the variability in rainfall  can lead to noise-induced transitions from a patterned state to the bare soil state  when the mean annual precipitation level puts the system in a bistable regime, below the transcritical bifurcation  point found in Section~\ref{sec:per:linstab}.
We show that both the rainfall and vegetation band characteristics can impact statistics of these collapse events.

We assume $N_s=2$ equal rainy seasons per year, each lasting $T_r$ days. The intervening dry seasons last for a time $T_d=T_y/N_s -T_r$, where $T_y=365$ $days$.  During each rainy season we model the rainstorms as a Poisson point process with a mean arrival rate of $\lambda_r={\cal MAP}/H_0/N_s/T_r$ where ${\cal MAP}$ is the mean annual precipitation and $H_0$ is the mean rainfall per storm.  The actual amount of rainfall $H_i$, in the $i$th storm, is drawn from an exponential distribution with mean $H_0$, where we typically consider $0.5\le H_0\le 2$, measured in $cm$. While  $T_r=1\ month$ is our default value, we let $T_r\to 0$ to speed up computations 
for the ecosystem collapse simulations of Section~\ref{sec:stoch:col}, after first exploring some of the effects of changing the rainy season duration in Section~\ref{sec:stoch:tr}. 

\begin{figure}
    \centering
    \includegraphics[width=\textwidth]{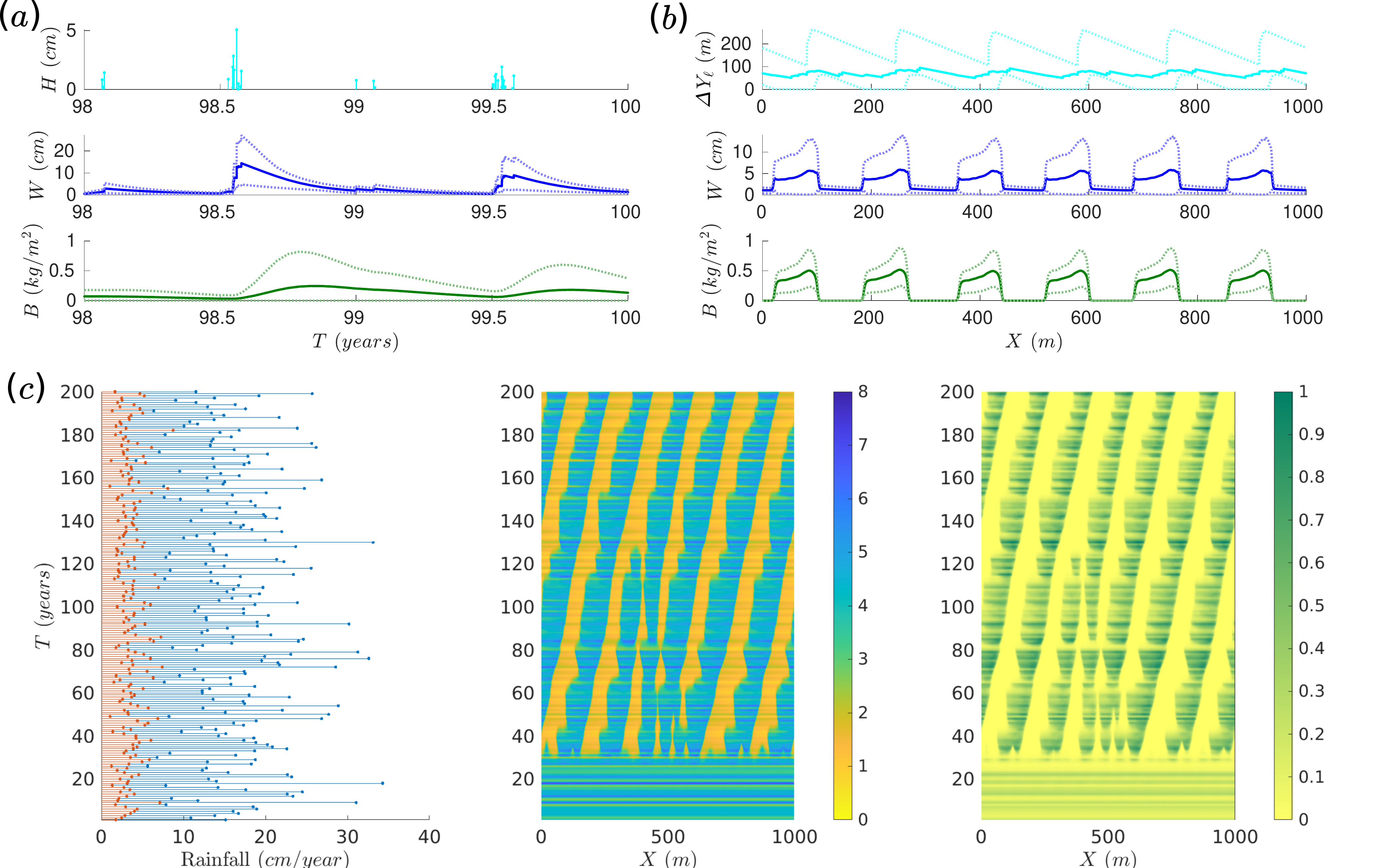}
    \caption{(a) Time series showing, from top to bottom, storm depth $H$ of rain pulses, domain-averaged soil water $W$ and biomass $B$ during last two years of a 100 year simulation under periodic rainfall. Solid lines indicate spatially averaged fields while dashed lines indicate instantaneous min/max. (b) Spatial distribution, from top to bottom, of dimensioned maximum distance $\Delta Y_\ell(X)$ traveled by water before infiltrating, soil water $W$ and biomass $B$ during last year. Solid lines indicate averages over rainstorms for $\Delta Y_\ell(x)$ and annual averages for $W$ and $B$.  Pointwise min/max values are shown by dashed lines.  (c) From left to right, time series of annual rainfall totals in blue with contribution from largest rainstorm highlighted in orange, spacetime plots of annually averaged soil water and biomass. Parameters: Stochatic rainfall with ${\cal MAP}=16\, cm$, mean storm depth of $H_0=1\, cm$ and $T_r=1\, month$ on a $L=1000\, m$ domain and initialized with 1\% random noise on top of the spatially uniform solution.   }
    \label{fig:nonlinStoch}
\end{figure}

Figure~\ref{fig:nonlinStoch} shows an example of results from a simulation on a $1\, km$ domain initialized with $0.1$\% random noise on top of the uniform vegetation state, ${\cal MAP}=16\, cm$ and mean storm depth $H_0=1\, cm$.  The simulation settles into a ``stochastic traveling wave'' solution which fluctuates from season to season, due to  rainfall variability, but can be characterized by a mean vegetation band width, spacing, and migration speed.  In this case the pattern consists of 6 bands on the domain and travels uphill on average, which is in contrast to the 6-band pattern obtained with periodic rainfall in Section~\ref{sec:per:nonlin}, which traveled downhill.  We also note the annual mean $\Delta Y_\ell$ in the last year, shown in Figure~\ref{fig:nonlinStoch}(b) has an average of $ 94\, m$, which is approximately half the wavelength of the pattern, which is $\sim 167\, m$. This ratio, at $\sim 0.56$, is remarkably in line with the resonance tongue phenomenon explored in Section~\ref{sec:per:linstab}.

\subsection{Comparison to periodic rainfall}

In repeated trials of the stochastic simulation, with parameters as in Figure~\ref{fig:nonlinStoch}, we   typically observe between 5 and 12 bands on the $1000\, m$ domain at 200 years.  This is in contrast to the skewed-lower and narrower range of 4 to 6 bands observed under periodic rainfall as described in Section~\ref{sec:per:nonlin}.  We  also see occasional, intermittent collapse of the vegetation to the bare soil state and explore this phenomenon further in Section~\ref{sec:stoch:col} at lower precipitation levels where the vegetation cannot recover. 

\begin{figure}
        \centering
        \includegraphics[width=.75\textwidth]{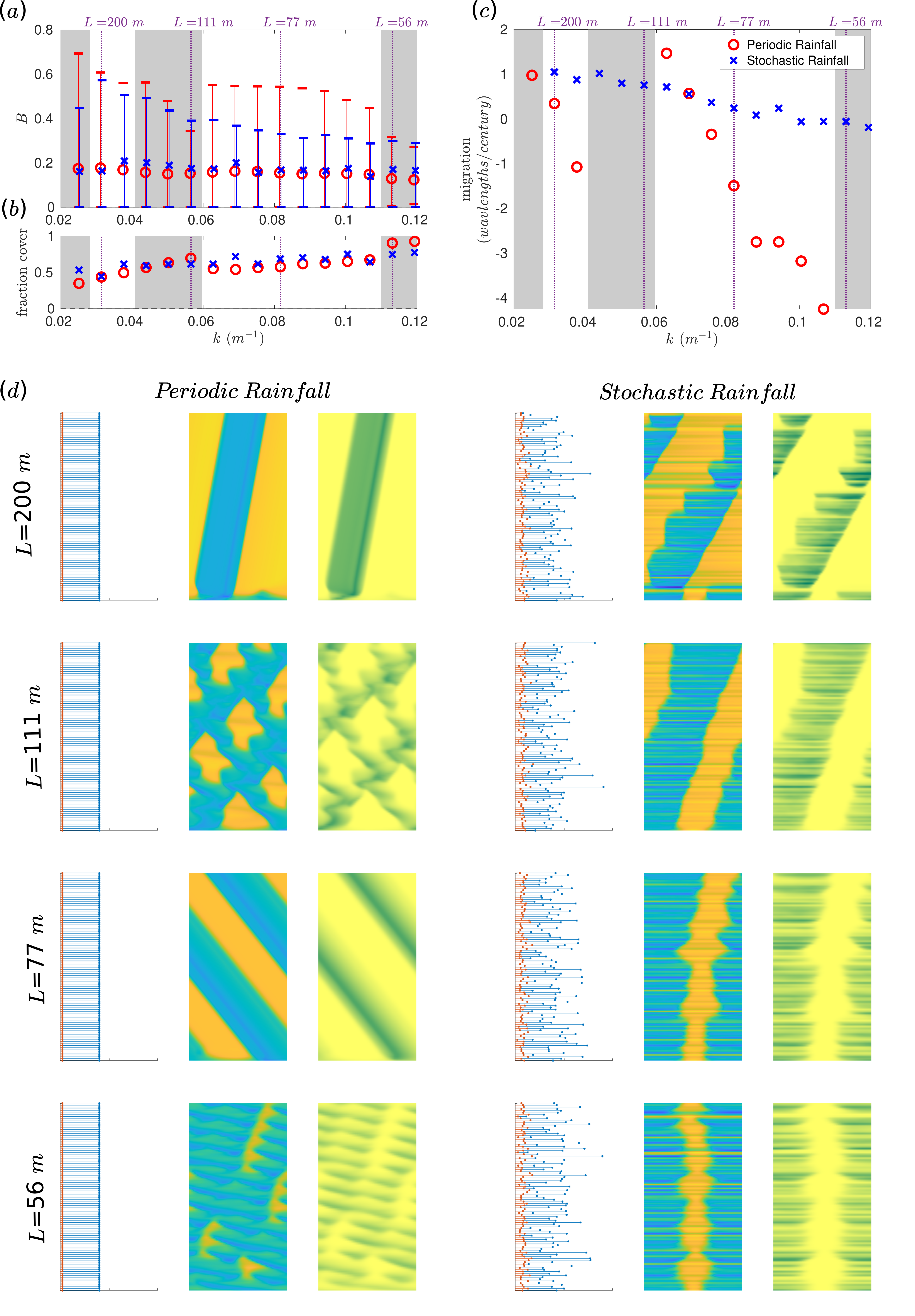}
        \caption{Comparison of pattern characteristics under periodic and stochastic rainfall. (a) Minimum, maximum and mean of annually averaged biomass profile, (b) fraction of domain covered by biomass and (c) mean migration speed, as a function of pattern wavenumber $k=2\pi/L$ for one band on a periodic domain of length $L$.  Results from simulations under periodic rainfall are marked with red circles, and stochastic rainfall with blue x's.  Either ``traveling waves" or ``stochastic traveling waves"  were not obtained from simulations with domain sizes appearing within the gray shaded regions.  Thumbnails of annual rainfall, soil water and biomass from example simulations are shown in panel (d) at the domain sizes indicated in the panels above for both (left) periodic rainfall and (right) stochastic rainfall.  The vertical time axis covers 100 years and the horizontal rainfall axis goes to 40 $cm$ in all of the cases, while each row has a different scale on the horizontal spatial axis for soil water and biomass in order to show the entire periodic domain.  Color scales are the same as in Figures~\ref{fig:nonlinPer} and~\ref{fig:nonlinStoch}.     }
        \label{fig:PerStochCompare}
    \end{figure}

Comparing  Figures~\ref{fig:nonlinPer} and~\ref{fig:nonlinStoch} shows that while fluctuations in the size and number of rainstorms per season drive fluctuations in the soil water-biomass dynamics, there are qualitative similarities between the mean pattern characteristics in the stochastic rainfall case and the periodic rainfall one.
However, Figure~\ref{fig:PerStochCompare}, which further compares the pattern characteristics of  periodic and stochastic rainfall, shows that this is not always the case.  To generate those results, we fixed the rain model parameters to have ${\cal MAP}=16\, cm$ and (mean) storm depth $H_0=1\, cm$, and enforced different band spacings by changing the size of the periodic domain $L$  and initializing with a perturbation of wavenumber $k=2\pi/L$.  We restricted to $L\le 250\, m$ since for domains with $L>250\, m$
($k<0.0252 \, m^{-1}$), 
the initial perturbation would split and the pattern eventually settled into a multi-band state on the domain. At $L=250\, m$, we obtained a single-band ``traveling wave" state for the periodic rainfall case, while the stochastic rainfall case still split into two bands.  For cases with $100 < L < 167$ $m$ and $L<59\, m$,  we did not observe ``traveling wave" states with periodic rainfall but did observe a ``stochastic traveling wave" state with stochastic rainfall. The intervals in $L$ for which one or both rainfall models did not reach a ``(stochastic) traveling wave state" are shaded gray in Figure~\ref{fig:PerStochCompare}(a-c).   We observed relatively consistent values of average biomass on the domain, but with a gradual increase in fraction of the domain covered as a function of $k$.  
Figure~\ref{fig:PerStochCompare}(c) shows migration frequency for these simulations. We compute the migration speed by tracking the motion of the uphill edge of the vegetation band, whenever there is a clearly defined band.   
We take the  edge as the location where the biomass first goes  above a threshold value of $\epsilon_B=Q/5=0.02 \, kg/m^2$.  While we see significant downhill migration with periodic rainfall, the stochastic simulation bands tend to travel slowly uphill (on average) except in a few cases with very short domain sizes.

\subsection{Dependence on rainy season duration}
\label{sec:stoch:tr}

In this subsection we investigate the impact of changing the duration of the rainy season on pattern characteristics, as well as on mean time to ecosystem collapse at low mean annual rainfall.
This investigation is  motivated in part by an additional computational speed-up that is possible if we take $T_r\to 0$ so that all  rainstorms in a given rainy season occur simultaneously.  In this limit we can use the same biomass profile for all the storms, thereby admitting an efficient parallel computation of the associated soil water contributions $\Omega$ by Equation~\eqref{eq:Omegax0}.  Exploring the impacts of rainy season duration is also of possible interest in light of  observed changes in the rainfall seasonality of Eastern Africa~\cite{wainwright2019eastern}.  

\begin{figure}
    \centering
    \includegraphics[width = \textwidth]{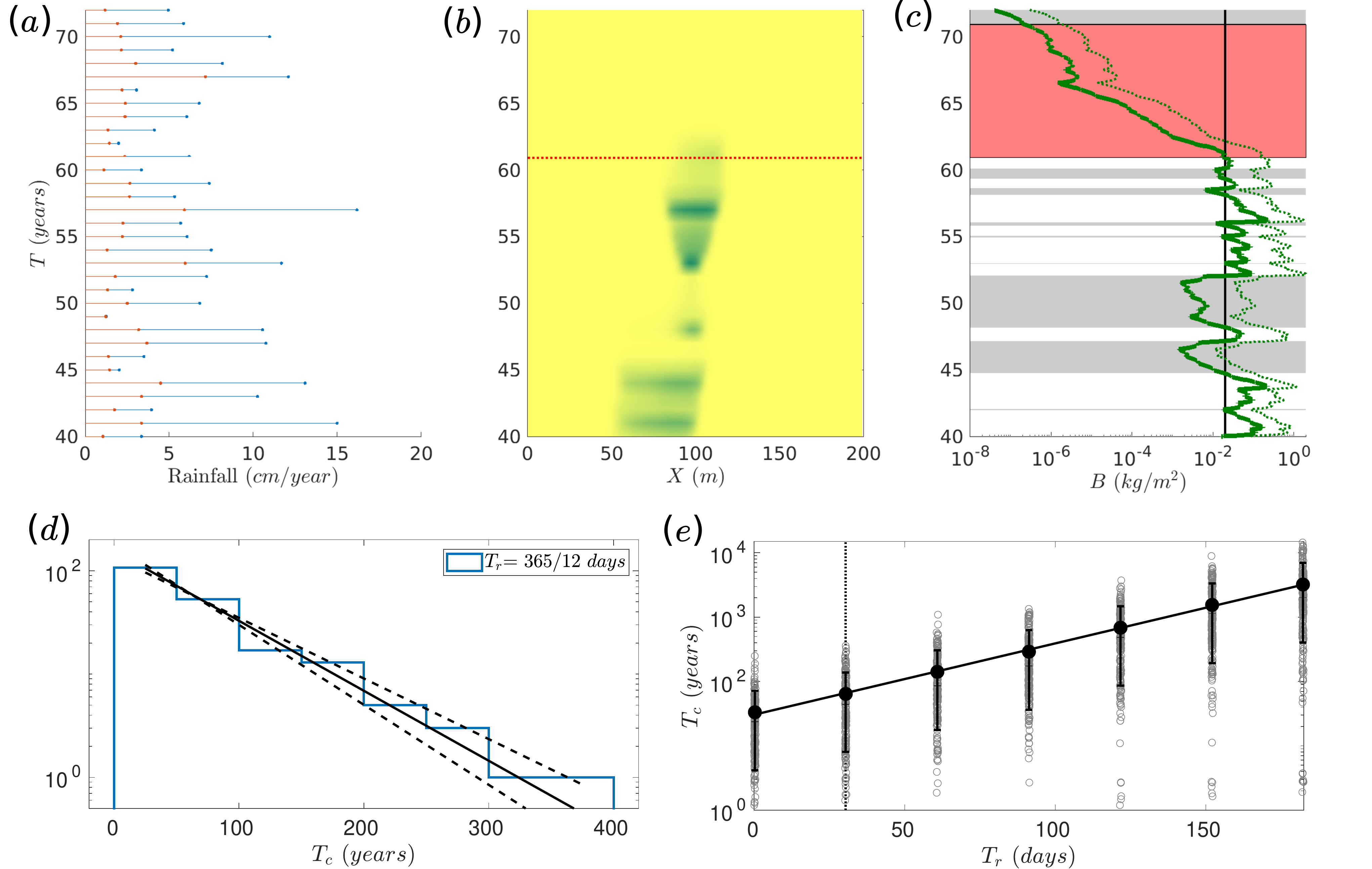}
    \caption{(a) Annual rainfall and (b) biomass distribution starting at $T=40$ $years$ for a stochastic rainfall simulation with ${\cal MAP}=8\, cm$, $H_0=1\,cm$ and $T_r=1\, month$ on a $L=200\, m$ domain that collapses at $T\approx 61\, years$. (c)  Collapse is defined when the mean biomass on the domain, indicated by the solid green line, falls below $\epsilon_B=0.02\, kg/m^2$ for 10 consecutive years (shaded red). All (shorter) intervals with $B_{avg}<\epsilon_B$ are shaded gray.  The dashed green line indicates the peak biomass value on the domain as a function of time.  (d) A histogram of collapse times from 200 trials with the parameters from (a)-(c), and indicated by a dotted vertical line in (e), has a mean survival time of $T_c=64\, years$.  The solid black line represents an exponential distribution with this mean, and the dashed lines represent a 95\% confidence interval for a maximum-likelihood fit of the histogram to an exponential distribution. (e) The mean survival time approximately follows an exponential trend as a function of the duration of the rainy season $T_r$. Each gray circle represents a single simulation, while the solid black   circles indicate the mean over 200 trials at each value of $T_r$, and the vertical bars indicate 95\% confidence interval for the fit of the trails $T_r$ to an exponential distribution. The data fits well the solid black line, $T_c= 30.5 \times 2.2^{T_r}$, with $T_r$ given in $months$ and $T_c$ given in $years$.       
    }
    \label{fig:train}
\end{figure}

Simulations at ${\cal MAP}=16\, cm$  indicate very little dependence of pattern characteristics on the duration of rainy season for $0<T_r\lesssim 3 \, months$, aside from an approximate $10-30\%$ increase in migration speed for each month of added rainy season duration.  
At a lower ${\cal MAP}$ value of $8\, cm$, where the bare soil state stably co-exists with patterns, simulations initialized with a banded vegetation state may transition to the stable bare soil state and not recover on  a centuries-long simulation timescale. In practice, we identify these collapse events by tracking the domain-averaged biomass; we use a threshold of $\epsilon_B=Q/5= 0.02\, kg/m^2$, and require the biomass level to stay below that for at least one decade.  Figures~\ref{fig:train}(a-c) illustrate a collapse event for a sample simulation at ${\cal MAP}=8\, cm$, $H_0=1\, cm$, and 1-month rainy season on a $L=200\, m$ domain.  Time intervals for which the average biomass, shown in Figure~\ref{fig:train}(c), falls below the threshold are shaded in gray, while the first interval that lasts 10 years is shaded red.  Survival times for 200 trials with the same parameters, shown in Figure~\ref{fig:train}(d), are  well-approximated by an exponential distribution. We use the same 200 rain sequences and rescale the relative wait times by a constant factor to obtain a range of rainy season durations between simultaneous rainstorms ($T_r\to 0$) and year-round rainfall ($T_r=6\,months$).  A  plot of the survival times as a function of duration of the rainy season, shown in Figure~\ref{fig:train}(e), indicates that the mean survival time  from the exponential fit
approximately doubles with each additional month of increase in rainy season duration.

\subsection{Collapse at low precipitation values}
\label{sec:stoch:col}

We now explore the dependence of collapse on rainstorm intensity and band spacing. 
We assume that all the rainstorms in each season occur simultaneously so that we may take advantage of the computational speed up noted in  Section~\ref{sec:stoch:tr} for $T_r\to 0$.  Since this may bias the collapse events towards shorter times, as suggested by results in Figure~\ref{fig:train}(e), we keep the focus on the trends associated with varying certain parameters.

We conduct trials with stochastic rainfall at ${\cal MAP}=8\, cm$, where stable ``traveling wave" patterns exist under periodic rainfall.  With stochastic rainfall, there is inevitably a collapse to the (bi-stable) bare soil in simulations initialized  with a single vegetation band on a domain of $50\le L\le 1000$ $m$ and mean rainstorm depth $0.4\le H_0\le 2$ $cm$.  Here, as illustrated in Figure~\ref{fig:train}(a-c), we define collapse to be when the domain-averaged biomass level first falls below $\epsilon_B=Q/5=0.02\, kg/m^2$ and remains so for a period of 10 years,  at which point we terminate the simulation.  We generate the initial condition for the stochastic simulations via a periodic rainfall simulation with an identical sequence of rainstorm depths in each season, run to its steady state. For this, rather than choosing each storm in the sequence to have the same strength, we take the expected number  of storms (rounded to nearest integer) that we will use for the stochastic model, and select the storm depths to match expected values for storm intervals of equal probability for the exponential distribution. We find that selecting the initial condition in this way avoids premature collapse due to the initial condition not being appropriately ``tuned'' to the rainfall pattern.

  \begin{figure}
        \centering
        \includegraphics[width=\textwidth]{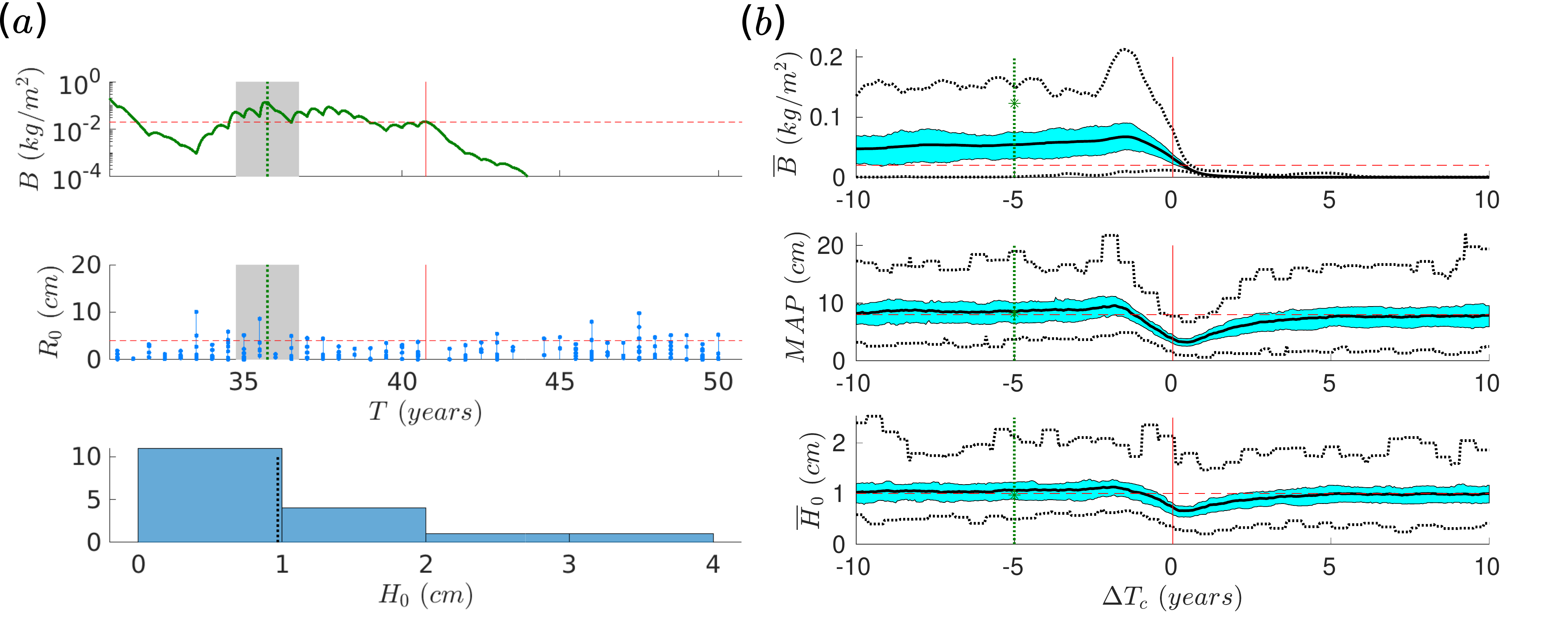}
        \caption{Rainfall statistics near collapse. (a) A time series of domain-averaged biomass is shown for a 20 year interval centered around a collapse event, marked by a vertical red line, is shown in the top panel.  The middle panel shows the seasonal rainfall, with the contribution from each individual storm marked by dots. The bottom panel shows a histogram of rainstorm depths within the 2-year window  highlighted in gray and centered at $\Delta T_c=-5$ years relative to the time of collapse. (b) From top to bottom, average biomass on the domain, mean annual precipitation, and mean storm depth 
        within a two-year window centered about the time relative to collapse, $\Delta T_c$.  The solid black line indicates averages over 200 trials, the shaded cyan region indicates the interquartile range, and the dotted lines indicate the minimum and maximum values.  The red dashed line indicates the collapse threshold level for biomass, and the expected mean values for the rainfall  parameters.
        \label{fig:RainfallCollapse}}
    \end{figure}

We begin with an exploration of rainfall statistics near collapse. Figure~\ref{fig:RainfallCollapse}(a) shows time series of the domain-averaged biomass and seasonal rainfall in the 20 year window surrounding collapse in a stochastic rainfall simulation with ${\cal MAP}=8\, cm$ and $H_0=1\, cm$ initialized with a single vegetation band on a 200 $m$ domain. The bottom panel of Figure~\ref{fig:RainfallCollapse}(a) shows a histogram of rainfall events within the shaded 2 year window centered at time $\Delta T_c=-5\, years$, which is measured relative to the collapse time $\Delta T_c=0$. We use this sliding 2-year window for computing rain statistics for the 200 trials. 
Figure~\ref{fig:RainfallCollapse}(b) then summarizes, from top to bottom, the mean biomass $\overline{B}$, the mean annual precipitation ${\cal MAP}$, and the mean rainstorm depth $\overline{H}_0$  for $\Delta T_c\in [-10,10]\, years$.  The average over all trials is given in solid black,  interquartile ranges are shaded cyan, and the minimum/maximum values are indicated by dotted lines.  The red dashed lines indicate the biomass threshold $\epsilon_B=0.02\; kg/m^2$ for collapse, and the expected values of ${\cal MAP}$ and $\overline{H}_0$ based on the parameters of the rainfall model. There is a noticeable drop in all three quantities in the vicinity of collapse, with biomass, by definition of collapse, failing to recover. These results suggest that the drop in ${\cal MAP}$ may be critical to driving collapse;  its minimum mean value falls from the expected 8 $cm$ to 3.2 $cm$; it does so approximately 5 $months$ after our defined ``collapse time".  The mean storm depth $\overline{H}_0$  also drops from its expected 1 $cm$ value to 0.66 $cm$. However,  this does not account for the full deficit in ${\cal MAP}$ since  we'd expect about 5.3 $cm$ if the mean storm frequency remained at 8 storms per year. The average ${\cal MAP}$, over the trials, falls to a lower value because there is a similar drop (not shown) in the mean number of storms per year, from 8  to approximately 5.

\begin{figure}
        \centering
        \includegraphics[width=\textwidth]{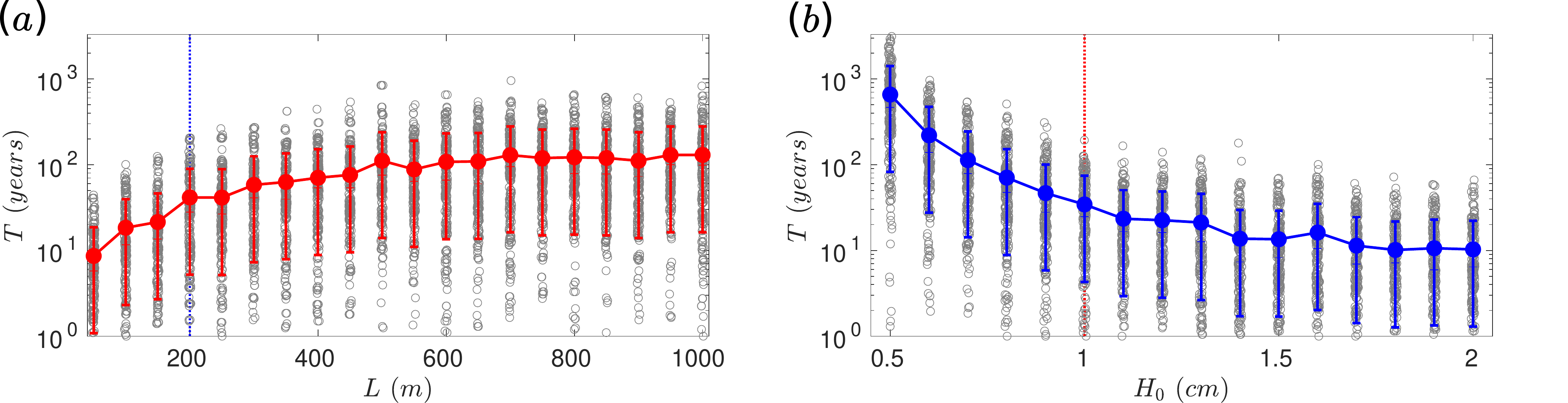}
        \caption{Survival times for simulations under stochastic rainfall with ${\cal MAP}=8\, cm$ and a single band initialized on the domain  as a function of (a) domain length $L$ with mean storm depth $H_0=1\, cm$ and (b) mean storm depth $H_0$ with domain length $L=200\, m$.  Each gray circle represents a single simulation, while the red/blue circles indicate the mean of a maximum-likelihood fit of the 200 trials at each parameter value to an exponential distribution, and the vertical bars indicate a 95\% confidence interval for each fit.  A linear interpolation of the mean values highlights the trend that survival times (a) increase as a function of domain size and (b) decrease as a function of mean storm depth.   
        }
        \label{fig:TcollapseLxH0}
    \end{figure}

Figure~\ref{fig:TcollapseLxH0}(a) shows the survival times, i.e. the length of time before collapse, of stochastic simulations with ${{\cal MAP}}=8\, cm$ and mean rainstorm depth $H_0= 1\, cm$ as a function of domain length $L$. 
The mean survival times from exponential fits of 200 trials at each value of $L$ are marked by solid red circles. An increase in the (periodic) domain size, corresponding to an increase in the spacing between bands in a periodic pattern, leads to longer mean survival time over this range.  However, the trend (on the logarithmic scale) appears to saturate for longer domain sizes.  Results analogous to those of Figure~\ref{fig:TcollapseLxH0}(a), described above, are shown with blue in Figure~\ref{fig:TcollapseLxH0}(b) as a function of the mean storm depth $H_0$ for a fixed domain size of $L=200\, m$.  We see longer survival times on average with smaller $H_0$, with the trend leveling off somewhere above $H_0\approx 1\, cm$. Since ${\cal MAP}$ is fixed, smaller storm depths correlate with more storms in each season and also less variability in annual precipitation from year to year.  

\begin{figure}
        \centering
        \includegraphics[width=\textwidth]{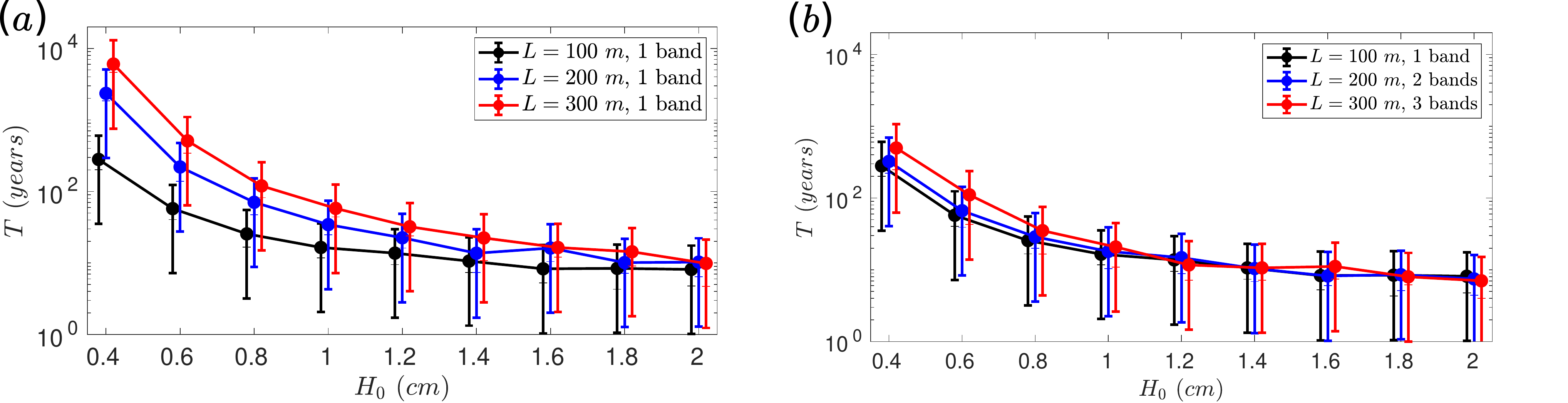}
        \caption{ How band spacing affects the dependence of survival times on mean storm depth $H_0$ for stochastic simulations with ${\cal MAP}=8\, cm$. (a) A single band is initialized on a $L=100,\, 200$ and $300\, m$ domain. (b) One, two and three bands are initialized with 100 $m$ spacing. The red, blue and black circles indicate the mean of a maximum-likelihood fit of the 200 trials to an exponential distribution at each parameter value, and the vertical bars indicate a 95\% confidence interval associated with each fit.  A linear interpolation of the mean values highlights the trends for these simulations. The mean survival times for simulations (a) initialized with a single vegetation band increase with domain size for mean storm depths below approximately $1\, cm$, yet (b) do not depend strongly  on domain size when the band spacing of the initial condition is fixed.}
        \label{fig:TcollapseLxNk}
    \end{figure}

We find that the mean survival time depends more strongly on $H_0$  for larger domain sizes when initialized with a single band on the domain.  Figure~\ref{fig:TcollapseLxNk}(a) shows the average survival times from a maximum-likelihood fit of 200 trials to an exponential distribution at each $H_0$ for $L=100$, 200, and $300\, m$ domains. 
We also observe that increasing the domain size but fixing the band spacing has little effect on trends in mean survival time.  Figure~\ref{fig:TcollapseLxNk}(b) shows results analogous to those of Figure~\ref{fig:TcollapseLxNk}(a) described above, except that the initial condition is chosen for each domain size to fix the band spacing at $100\, m$.  Some of the longer-lived trials initialized with multiple vegetation bands do occasionally prolong survival by first loosing one or two bands. However, these partial collapse events are infrequent and have little impact on the overall statistics;  the majority of simulations collapse by losing all vegetation bands at once.    

\section{Discussion}
\label{sec:dis}

We have developed a pulsed-precipitation model for  banded vegetation patterns in dryland ecosystems and used it to investigate the impacts of changing rainfall patterns.  The model is built upon the fast-slow modeling framework~\cite{gandhi2020fast}, and leverages additional simplifying  assumptions about overland surface water flow and infiltration into the soil to obtain a closed-form expression for the soil water contribution from a rain storm.  Biomass and soil water evolve on the slow timescale associated with plant growth, with rain storms modeled as instantaneous kicks to the soil water,  which are spatially dependent as they follow the biomass profile.  These soil water kicks capture the positive biomass-water resource feedbacks via enhanced infiltration and reduced surface flow speeds in vegetated zones.    

Our pulsed-precipitation model paves the way for exploration of stochastic rainfall patterns by allowing significant computational speed-up over the original fast-slow model, thus making large numbers of trials feasible.   We note that computational speed has also been addressed by applying machine learning techniques to predict the soil water distribution following rain in a more detailed hydrological model~\cite{crompton2021sensitivity}.  Although stochasticity and seasonality of rainfall were not considered in that paper, they did include storm duration, along with storm depth, as  training parameters, and observed similar qualitative trends of increased  band spacing with increased storm depth. An advantage of our approach is that simplifying the model keeps analysis, and the insights gained from it, within reach. Nonetheless, it will be important to characterize the impacts of our simplifications on predictions in future work through  comparison to more detailed models.  

Linear stability analysis of the model under periodic rainfall reveals that the distance $\ell_0$ that water flows on the surface before infiltrating into the soil plays a key role in determining pattern characteristics such as band spacing, a result that has also been suggested in the context of so-called ``flat-terrain" vegetation patterns~\cite{thompson2011vegetation}.  With periodic rainfall, the pattern formation in the pulsed-precipitation model is organized around a series of ``spatial resonances" in which water from the newly-forming bare soil region travels some integer number of wavelengths of the pattern before infiltrating into the newly-forming vegetation band. Simulations indicate that while the nonlinear patterns that form are significantly different from those predicted by the linear stability analysis, insights about the key role of the distance surface water flows still apply, even under stochastic rainfall. Nonetheless, some of the predictions obtained under periodic rainfall, such as significant downhill migration of vegetation bands, run contrary to observation.  This aspect of the underlying resonance structure, present with the idealized periodic rainfall, is however washed out when we introduce variability to the rainfall model; stochastic rainfall simulations produce banded patterns with characteristics that are reasonably consistent with observation. The impact of stochasticity on the existence and stability of nonlinear traveling wave patterns is itself an intriguing mathematical question. For example, how does the so-called Busse balloon, which was investigated in the context of vegetation patterns for a modified Klausmeier model~\cite{van2013rise,bastiaansen2018multistability}, change when rainfall is less predictable? Our investigation of a stochastic and impulsively forced  pattern forming system suggests new directions for fundamental pattern formation research.  

Motivated by the potential for identifying precursors to ecosystem collapse in a changing climate, a main focus of this study is the transition from  spatially-patterned vegetation to the desert state.  Indeed, at low enough mean annual precipitation values, stable vegetation patterns exist alongside a stable bare soil state, and fluctuations in rainfall can trigger ecosystem collapse. We see that both the pattern characteristics, such as band spacing, and the rainfall characteristics, such as rainy season duration and mean storm depth, have an impact on the mean time to collapse. Increased band spacing, corresponding to a larger area for harvesting water, leads to longer survival times. Both longer rainy seasons, corresponding to shorter dry intervals in which the biomass must survive without rainfall, and less intense storms, corresponding to a decrease in variability of rainfall from season to season, also increases the mean survival time.
We note that we have also observed collapse of patterns at higher precipitation levels but, if the bare soil state is unstable,  the ecosystem is expected to recover.

     \begin{figure}
        \centering
        \includegraphics[width=\textwidth]{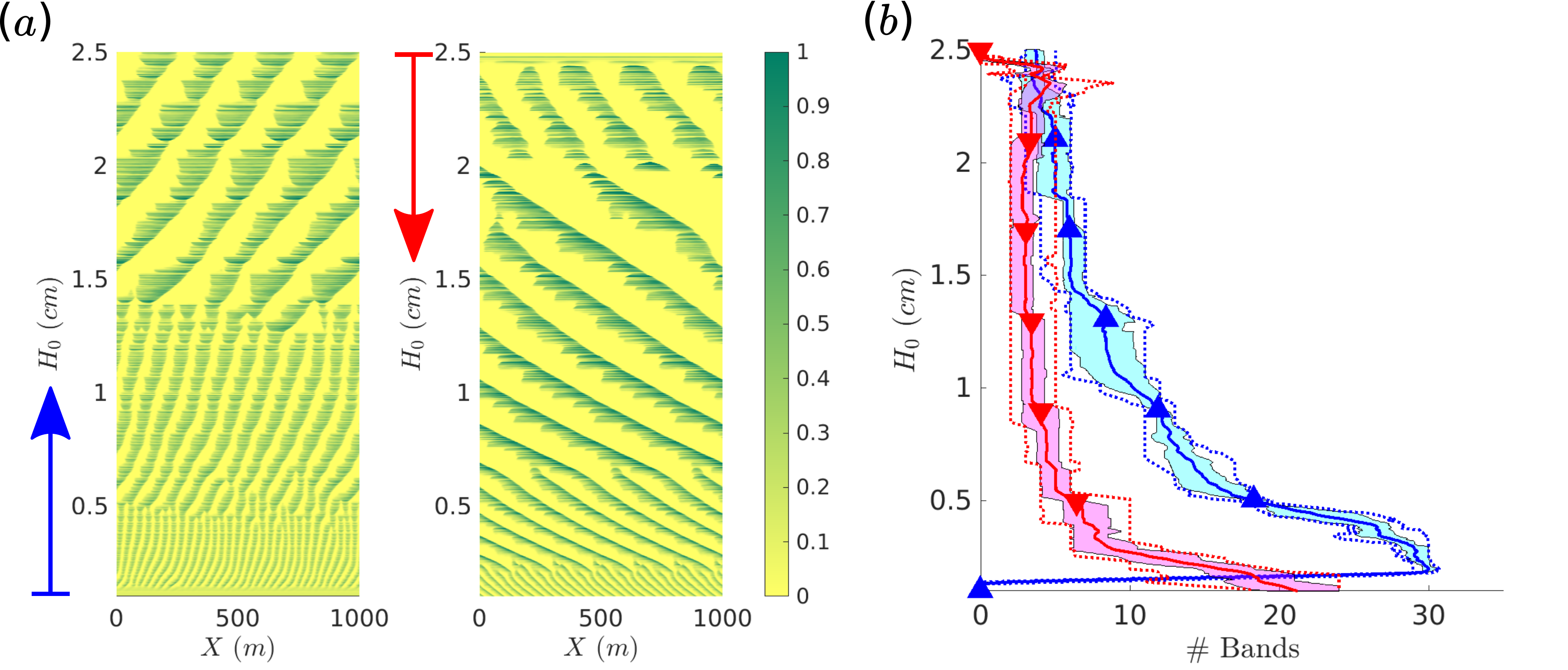}
        \caption{Transitions in band spacing as a function of slowly varying mean storm depth $H_0$.  (a) Spacetime plots of annually averaged biomass distribution from 1200-year simulations with stochastic rainfall with ${\cal MAP}=16$ $cm$ and rainy season length $T_r\to 0$. The mean storm depth is linearly increased from $H_0=0.1$ $cm$ to $2.5$ $cm$ at a rate of $2$ $cm/millennium$ in the left panel and decreased at the same rate starting from $H_0=2.5$ $cm$ in the right panel.  The blue and red arrows indicate the direction of time for increasing and decreasing storm depth.  (b) The average number of bands from 200 trials of increasing and decreasing $H_0$ are depicting with the think blue and red lines marked by up and down triangles.  The shaded cyan and magenta regions indicate the interquartile range from the 200 trails at each value of $H_0$ for increasing and decreasing $H_0$. The dotted lines indicate the maximum and minimum number of bands from the trials for each corresponding color.             
        }
        \label{fig:StochH0ramp}
    \end{figure}
Studies that use mathematical models to investigate the possible impact of a changing climate on vegetation patterns typically do so by varying the mean annual precipitation~\cite{siteur2014beyond,siero2015striped,bastiaansen2020effect}, with all other parameters held fixed. The pulsed-precipitation model allows for exploration of  other rainfall characteristics, such as mean storm intensity, seasonality and other forms of variability. We illustrate the potential here by presenting results from 200 trials with stochastic rainfall at $\mathcal{MAP}=16\, cm$ and $T_r\to 0$ in which the mean storm depth starts at $H_0=0.2\, cm$, and is then slowly increased to $H_0=2.5\, cm$  (at a rate of $2\, mm/century$).  We also carry out another 200 trials with the mean storm depth slowly decreasing from $H_0=2.5\, cm$, at the same slow rate. (Each of these 400 trials is initialized with uniform vegetation with $1\%$ random noise.)  Figure~\ref{fig:StochH0ramp}(a) shows  example spacetime distributions of the annually-averaged biomass, where the blue and red arrows indicate the direction of time for the simulations with increasing and decreasing storm intensity.  As expected, based on the role of storm depth in  pattern selection for the model, the simulations exhibit increased band spacing, on average, at higher $H_0$.  Figure~\ref{fig:StochH0ramp}(b) summarizes the number of bands on the domain as a function of $H_0$ from all of the trials for increasing $H_0$ (blue) and decreasing $H_0$ (red).  We note that band merging and band splitting events were also observed in a study of the extended Klausmeier model when the annual mean precipitation was slowly ramped down and back up~\cite{siteur2014beyond}.  
Those typically occurred as spatial period-doubling or period-halving in this simpler deterministic model setting. In the stochastic pulsed-precipitation model results shown in Figure~\ref{fig:StochH0ramp} there is more variability in the band loss and gain events, which here occur without any changes in the mean annual precipitation.

We have made a number of simplifying assumptions in this work,  particularly in the overland flow model, with the goal of allowing for analytic insight and computational efficiency. The form of the infiltration model neglects soil saturation effects, which may become important when considering very intense rainstorms.
The choice of periodic boundary conditions does not allow for exploration of the impact of surface water run off. Both effects have been considered in~\cite{siteur2014will,crompton2021sensitivity} using very different approaches to model surface hydrology. 
We note that the work of Crompton and
Thompson~\cite{crompton2021sensitivity}  indicates that storm duration plays an equally important role as storm depth in pattern formation, as both can affect rainfall intensity. Our assumption of instantaneous rain pulses does not allow us to explicitly explore the impact of storm duration, separate from storm depth, on pattern formation. The possibility that we might reinterpret our 
storm depth parameter, which determined pattern wavelength in our pulsed model, as capturing an effective surface water height during storms will be the subject of future work, in which storm duration is included in our model.

Our focus in this work has been on capturing the influence of hydrological processes across timescales, and the biomass model used here is based directly on previous conceptual models~\cite{rietkerk2002self,gilad2004ecosystem}.  Other works have explored the impacts of incorporating additional vegetation characteristics and processes~\cite{pueyo2008dispersal,bennett2019long,eigentler2020effects,saco2007eco,guttal2007self,dordorico2006vegetation,borgogno2007effect}. Fortunately, as indicated in Appendix A, we did not find a strong dependence of simulation results on the biomass diffusion rate, which is a phenomenological parameter that is not well-constrained by observation.

Generalizing the pulsed precipitation framework to capture the influence of heterogeneous terrain and moving to two spatial dimensions would open the door to a number of possible future directions.  For example, an investigation of the Klausmeier model~\cite{Klausmeier1999} with topographically modified water transport suggested that the placement of patterns relative to local valleys and ridges may provide an indicator for resilience of the ecosystem to drought~\cite{gandhi2018topographic}. It would be interesting to explore what additional insights could be gained by a two-dimensional pulsed-precipitation model that captures the influence of various rainfall characteristics, not just the mean precipitation value that controls drought.  Capturing hydrology on the fast timescale could also allow for the exploration  of the impact of roads, noted for example in~\cite{gowda2018signatures}, or other disruptions to surface water flow on the vegetation patterns.  It is also likely to be important when coupling to landscape evolution through erosion and sediment transport. 

Theoretical studies have suggested that spatial patterns can increase ecosystem resilience, and protect it against collapse under a decrease in total rainfall~\cite{mau2015reversing,rietkerk2021evasion}.  However, climate change will impact not only the  yearly  mean rainfall. It is already seen to disrupt  seasonality in rainfall patterns, and increase variability in storm characteristics.  A framework like the pulsed-precipitation model, which can capture the influence of changing rainfall patterns, may therefore be useful since it can assess resilience in those contexts.   
It also brings into sharp focus the driving role of the fast hydrological processes on the dynamics of dryland vegetation patterns, which then evolve on their own years-to-decades timescales. This highlights the potential for time-resolved data from field-based hydrology monitoring, across vegetation bands, as a welcome and timely feedback to mathematical modeling efforts.

\vskip6pt

\enlargethispage{20pt}







\paragraph{Acknowledgements.} We would like to thank Merlin Pelz, Justin Finkel, Shiv Agrawal, Karna Gowda and Sarah Iams for helpful discussions related to this work.  
Computing resources were provided by the High Performance Research Computing (HPRC) Core Facility at Virginia Commonwealth University. This work of MS was supported in part by National Science Foundation grants DMS-1517416 and DMS-2023109.
\bibliographystyle{unsrt}
\bibliography{references}

\appendix

\section{Dependence of pattern formation on soil water and biomass diffusion rates}
\label{appendix:DBW}
\begin{figure}
        \centering
        \includegraphics[width=0.9\textwidth]{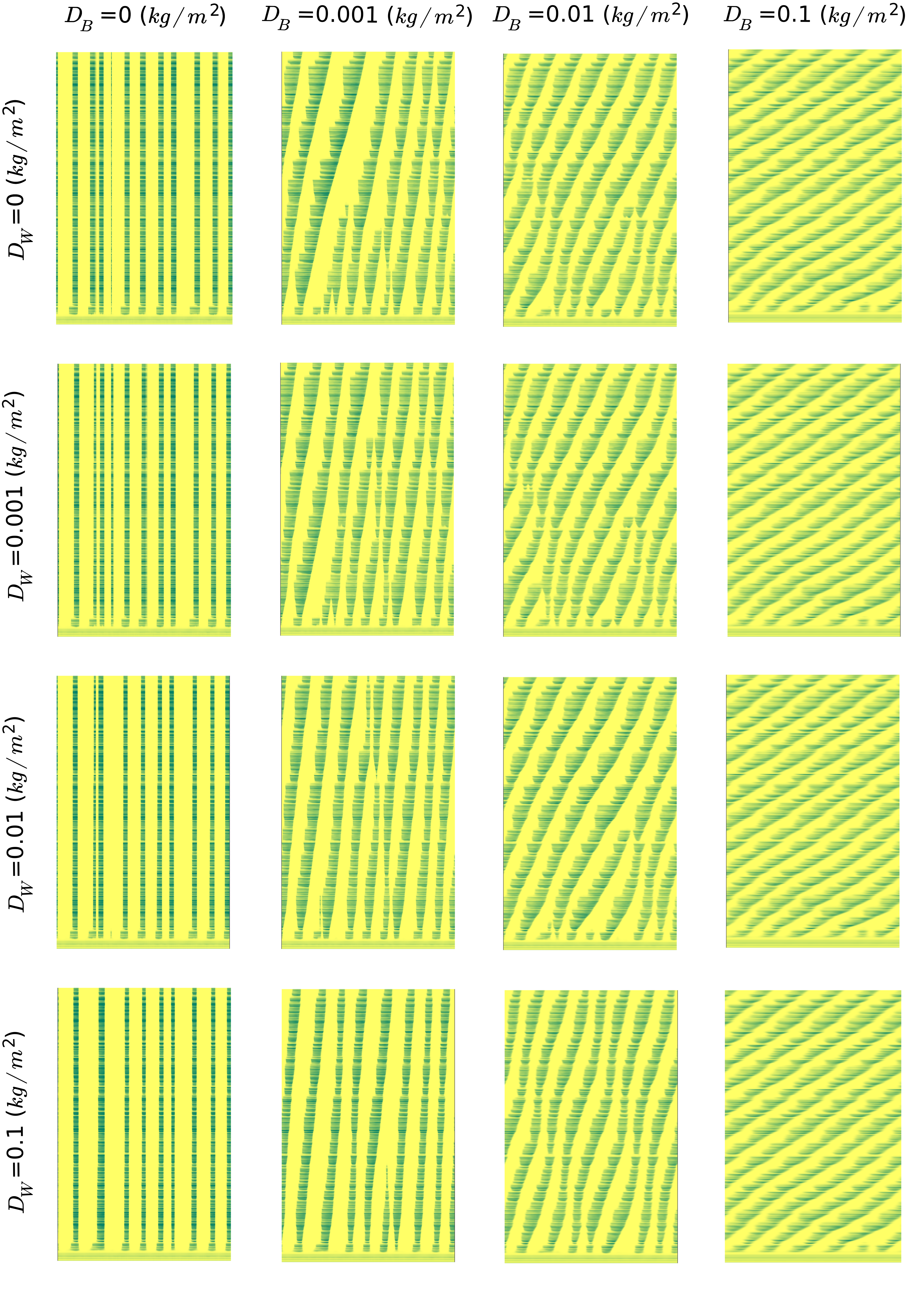}
        \caption{Dependence of pattern characteristics on biomass and soil water diffusion rates. Thumbnail plots of annually averaged  biomass in units of $kg/m^2$ as a function of space (horizontal axis) and time (vertical axis) are shown for different  biomass and soil water diffusion rates $D_W,\, D_B=0,\, 0.001,\, 0.01$ and $0.1\, kg/m^2$. The color scale is the same as in Figures~\ref{fig:nonlinPer} and~\ref{fig:nonlinStoch}. Each simulation has the same stochastic rainfall sequence for  a total of $500\, years$, with ${\cal MAP}=16\, cm$, mean storm depth $H_0=1\, cm$ and rainy season duration $T_r\to 0$ on a $L=1000\, m$ domain, and initialized with the same 1\% random noise on top of the spatially uniform solution.
        }
        \label{fig:appendix:DBW}
    \end{figure}

The soil water and biomass diffusion rates $D_W$ and $D_B$ are typically not well-constrained by observation in reaction-diffusion models of vegetation pattern formation (See, e.g.~\cite{gandhi2019vegetation}). This appendix explores the dependence of pattern formation on these constants in the pulsed-precipitation model with stochastic rain input.  We take $\mathcal{MAP}=16\, cm$, mean storm depth $H_0=1\,cm$ and $T_r\to 0$ as is done in Section~\ref{sec:stoch:col}. Five-hundred year simulations with the same initial condition and rainfall sequence but different diffusion rates are carried out on a 1000-$m$ domain. The thumbnails of spacetime plots of biomass shown in Figure~\ref{fig:appendix:DBW} illustrate the qualitative influence of $D_B$ on the uphill migration rate of the pattern. We see relatively minor impact from changing $D_W$ across three orders of magnitude.  Importantly, selection of the pattern wavelength is insensitive to the values of \textit{both} diffusion constants.

\section{Comparison of pulsed-precipitation model to fast-slow model}\label{appendix:CompareFastSlow}
 \begin{figure}
        \centering
        \includegraphics[width=0.9\textwidth]{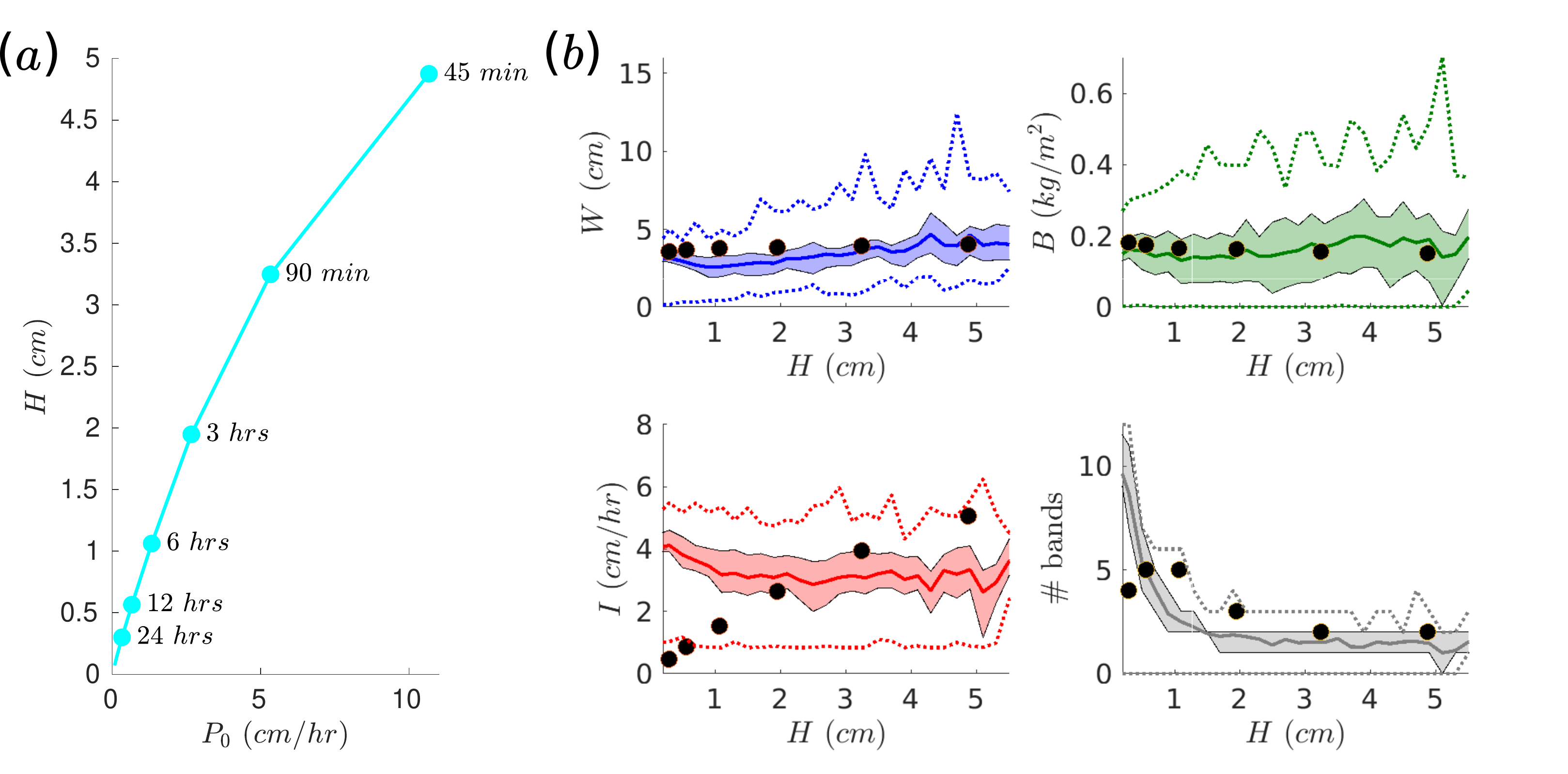}
        \caption{A comparison of the pulsed-precipitation model to the fast-slow model~\cite{gandhi2020fast}. Simulations of the fast-slow model  use a fixed mean annual precipitation $\mathcal{MAP}=$ 16 $cm$ over two equally spaced storms per year with durations $T_{dur}=45\, min$ up to  $24\, hrs$, and are initialized with 1\% noise on top of a uniform state and a domain of $L=500$ $m$.  The results of 200-year simulations are compared to analogous results from 150 trials of the pulsed precipitation model with different stochastic rainfall with mean storm depths from $H= 0.2$ $cm$ up to $6$ $cm$.   (a) The domain-averaged, temporal-maximum surface water height from simulations of the fast-slow model during the storm is shown as a function of rainfall rate, and the points are labeled by the storm duration. (b) The domain-averaged soil water, biomass, infiltration rate and number of bands from the final year of the fast-slow simulations, indicated by black circles are compared to analogous results from the stochastic pulsed-precipitation model.  In each case the mean value of the quantity as a function of the mean storm depth $H$ is plotted with a thick solid line.  The interquartile range is shaded, and the minimum/maximum vlaues are indicated by dotted lines.   }
        \label{fig:appendix:CompareFastSlow}
    \end{figure}
This appendix provides a comparison of simulation results from the fast-slow model~\cite{gandhi2020fast}, described in Section~\ref{sec:fastslow}, to corresponding results from the pulsed precipitation model, presented in Section~\ref{sec:pulsedprecip}. Some care must be taken to make the comparison as the fast-slow model incorporates both storm depth and duration, whereas the pulsed precipitation model assumes instantaneous rain impulses (with no duration).  We can, however, account for this by interpreting the storm depth associated with an impulsive rain storm from the pulsed-precipitation model as an effective surface water height that is achieved during a storm of finite duration in the fast-slow model.   

Figure~\ref{fig:appendix:CompareFastSlow} shows results from simulations of the fast-slow model using the same parameters as~\cite{gandhi2020fast}, which are also reported in Section~\ref{sec:fastslow}.  A mean annual rainfall of $\mathcal{MAP}=16\, cm$ is used, and each of two identical rainy seasons per year is modeled by a single storm with constant rainfall rate $P_0$ of duration $T_{dur}= 45\, min$, $90\, min$, $3\, hrs$, $6\, hrs$, $12\, hrs$ and $24\, hrs$. Simulations with each of these six rain storm durations are initialized with 1\% random noise on top of a uniform vegetation state and run for 200 years.  We take the peak surface water height achieved at each point on the 500 $m$ domain during the final cycle of the fast system (i.e. the time period over which water from the last rainstorm of the simulation remains on the surface), and plot the spatial average of this profile (denoted by $H$) as a function of the rainfall rate during the storm in Figure~\ref{fig:appendix:CompareFastSlow}(a).  

This domain-averaged peak surface water height $H$, captured by Figure~\ref{fig:appendix:CompareFastSlow}(a), 
provides a path to compare the fast-slow model to the pulsed-precipitation model.  In particular, we interpret the mean storm depth parameter $H_0$ of the pulsed-precipitation model as an effective peak surface water height, analogous to the quantity $H$ from the fast-slow model described above.  Instead of taking $H_0$ as the total mean rainfall during storms, we can then think of it as characterizing an effective peak surface water height during storms that takes into account both the influence of storm depth and duration.  We run 150 total trials of the stochastic pulsed-precipitation model, also with $\mathcal{MAP}=16\, cm$ and two rainy seasons per year on a 500 $m$ domain initialized with 1\% noise added to a uniform vegetation state. We do five trails at each mean storm depth value  $0.2\, cm\, \le H_0\le 6 \, cm$ at 0.2 $cm$ increments. For the last twenty years of each 200-year trial, we compute the mean storm depth during that year, the number of bands in the pattern, as well as the domain-averaged peak biomass and soil water during the year, and the domain averaged infiltration rate during the storm pulse.  These are compared to the analogous quantities obtained with the fast-slow model in Figure~\ref{fig:appendix:CompareFastSlow}(b).  

In each of the four panels shown in Figure~\ref{fig:appendix:CompareFastSlow}(b) the mean value of the quantity, as a function of mean storm depth from the stochastic pulsed-precipitation simulations, is plotted with a solid line.  The interquartile range is shaded and the minimum and maximum values obtained from the trials is also indicated by dotted lines.  The black circles indicate values from the final year of the fast-slow simulation.   We see agreement in the biomass and soil water levels between the two models.  The number of bands is also consistent between the two models for larger $H$.  At small $H$, the difference in the infiltration rate functions between the two models may explain the differences in the predicted band spacing.  Indeed, we see improved agreement in both lower panels of Figure~\ref{fig:appendix:CompareFastSlow}(b) in simulations (not shown) where we decrease the parameter $A$ in the infiltration function of the fast-slow model below it's default value $A=1\, cm$. Specifically, we refer to the the factor $H/(H+A)$ in the fast-slow infiltration function given in Table~\ref{tab:infadv}; decreasing $A$ makes the fast-slower infiltration closer to the step-function used in the pulsed-precipitation model. We note that Thompson et al.~\cite{thompson2011vegetation} have explored the dependence of the infiltration rate on surface water height in the context of so-called ``flat-terrain" vegetation patterns.   

Lastly, we emphasize that the computational savings of the pulsed-precipitation model over the fast-slow model are significant.  We see a factor of 200 or more speed up in simulation time by going from the fast-slow model to the analogous pulsed-precipitation simulation in the comparisons presented here.

\section{Linear stability of bare soil state to uniform perturbations}
\label{app:transcrit}
\begin{figure}
        \centering
        \includegraphics[width=\textwidth]{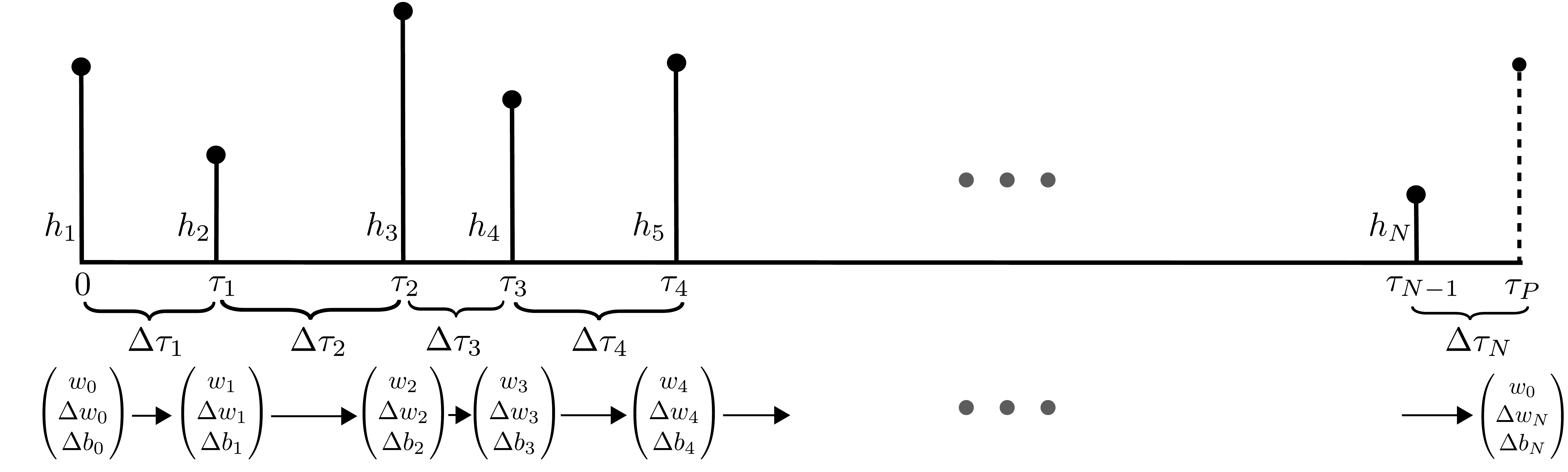}
        \caption{Summary of rainfall model and other quantities introduced for the linear stability analysis.}
        \label{fig:appendix}
    \end{figure}

In this appendix we consider the linear stability of the zero biomass desert state to spatially uniform perturbations. (We need not consider heterogeneous perturbations, proportional to $e^{ikx}$, for the zero biomass state since we require $b\ge 0$.) We show that the ${\cal MAP}$
threshold for loss of stability, denoted ${\cal MAP}_c$,  is independent of details of the rainfall model. 
We consider $N$ pulses of rain per year, with strengths $h_k$, $k=1,\ldots, N$, which repeat annually and sum to ${\cal MAP}$. The temporal spacing between pulses is denoted $\Delta\tau_k$, $k=1,\ldots,N$, and these time intervals sum to $\tau_P=3.65$, i.e. one year in dimensionless units. The periodic rainfall model, and the parameters needed for the linear stability analysis is summarized by Figure~\ref{fig:appendix}.

The base state for the linear stability analysis has $b=0$ and a soil moisture level that repeats with the annual period.  Let $w_0$ be the initial condition for this periodic base state, and denote the soil moisture at time $\tau_k$, prior to the $k^{th}$ pulse of strength $h_k$, by $w_k$.
From \eqref{eq:psihk}, we have that a pulse of strength $h$ adds $\alpha h$ to $w$, and, from \eqref{eq:slow:odes:linear:k:w}, that a dry period of duration $\Delta\tau$ leads to its evaporative decay by a factor $e^{-\sigma\Delta t}$. It follows that
\begin{equation}
\label{eq:wkplus1}
w_{k+1}=(w_k+\alpha h_1)e^{-\sigma\Delta t_k},\quad k=0,\ldots, N-1.
\end{equation}
Moreover, we require $w_N=w_0$ for the periodic state, which determines $w_0$. 

We perturb this $(w,b)=(w(\tau),0)$ base state by $(\Delta w_0,\Delta b_0),$ at time $\tau=0$. This perturbation  advances to a value $(\Delta w_1,\Delta b_1)$ at time $\Delta \tau_1$, and so on. We obtain, for example, $(\Delta w_1,\Delta b_1)$ by evolving the following linearized slow system equations for time $\Delta \tau_1$
\begin{equation*}
\frac{d}{d\tau}\begin{pmatrix}
{\Delta w}\\
{\Delta b}
\end{pmatrix}=\begin{pmatrix}
-\sigma&\gamma w(\tau)\\
0& w(\tau)-1
\end{pmatrix}\begin{pmatrix}
{\Delta w}\\
{\Delta b}
\end{pmatrix},
\end{equation*}
 where here $w(\tau)=(w_0+\alpha h_1)e^{-\sigma \tau}$, for $\tau\in(0,\Delta\tau_1)$. From this we obtain the following map
\begin{equation*}
    \begin{pmatrix}
{\Delta w_1}\\
{\Delta b_1}
\end{pmatrix}=\begin{pmatrix}
e^{-\sigma \Delta \tau_1}&*\\
0&e^{\chi_1}
\end{pmatrix}\begin{pmatrix}
{\Delta w_0}\\
{\Delta b_0}
\end{pmatrix},
\end{equation*}
where the off-diagonal term $*$ is not needed for determining stability, and 
\begin{equation*}
    \chi_1=
\int_0^{\Delta \tau_1}(w(\tau)-1)\ d\tau=\Bigl(\frac{w_0+\alpha h_1}{\sigma}\Bigr)\Bigl(1-e^{-\sigma \Delta \tau_1}\Bigr)-\Delta \tau_1=\Bigl(\frac{w_0+\alpha h_1-w_1}{\sigma}\Bigr)-\Delta \tau.
\end{equation*}
Here the final equality follows from \eqref{eq:wkplus1}.
Repeating this for the $N$ rain pulses,
 we find
\begin{equation*}
    \begin{pmatrix}
{\Delta w_N}\\
{\Delta b_N}
\end{pmatrix}=\begin{pmatrix}
e^{-\sigma \tau_P}&*\\
0&e^{\chi_1+\chi_2+\cdots+\chi_N}
\end{pmatrix}\begin{pmatrix}
{\Delta w_0}\\
{\Delta b_0}
\end{pmatrix},
\end{equation*}
where
\begin{equation*}
\chi_k=\int_{\tau_{k-1}}^{ \tau_k}(w(\tau)-1)\ d\tau=\Bigl(\frac{w_{k-1}+\alpha h_k-w_k}{\sigma}\Bigr)-\Delta \tau_k.
\end{equation*}
The stability boundary, denoted ${\cal MAP}_c$ is determined by the condition $\chi_1+\cdots+ \chi_N=0$. Using the fact that $w_0=w_N$ and that $\Delta\tau_1+\cdots+\Delta\tau_N=\tau_P$, it follows that 
\begin{equation*}
{\cal MAP}_c= h_1+\cdots+h_N=\frac{\sigma\tau_P}{\alpha},
\end{equation*}
which, in dimensioned quantities, is
$(LM/C\Gamma)365=10.95\ cm/yr$ for the parameters of Table~\ref{tab:dim}.

\end{document}